%% file: Paper_0lepton.tex
\pdfoutput=1
\newcommand*{\ATLASLATEXPATH}{}
\documentclass[cernpreprint,texlive=2011,txfonts,UKenglish]{\ATLASLATEXPATH atlasdoc}

\usepackage[subfigure,block=none]{\ATLASLATEXPATH atlaspackage}

\usepackage{multirow}
\usepackage{bm}
\usepackage{slashed}
\usepackage{mathrsfs}
\usepackage{xspace} %
\usepackage{units}
\usepackage{rotating}
\usepackage{verbatim}
\usepackage{float}
\usepackage{hyperref}      
\usepackage{cleveref}

\usepackage{\ATLASLATEXPATH atlasbiblatex}

\usepackage{\ATLASLATEXPATH atlascontribute}

\usepackage{\ATLASLATEXPATH atlasphysics}

\usepackage{\ATLASLATEXPATH atlasbsm}

\graphicspath{{logos/}{figures/}}

\usepackage{Paper_0lepton-defs}

\hypersetup{pdftitle={ATLAS draft},pdfauthor={The ATLAS Collaboration}}

\AtlasTitle{Search for squarks and gluinos in final states with jets and missing transverse momentum at $\sqrt{s}$ =13 \TeV\ with the ATLAS detector}

\AtlasRefCode{SUSY-2015-06}

\PreprintIdNumber{CERN-EP-2016-080}

\AtlasJournalRef{\EPJC (2016) 76: 392}
\AtlasDOI{10.1140/epjc/s10052-016-4184-8}

\AtlasAbstract{%
A search for squarks and gluinos in final states containing hadronic jets, missing
transverse momentum but no electrons or muons is presented. 
The data were recorded in 2015 by the ATLAS experiment in $\sqrt{s}=13\TeV$ proton--proton collisions at the Large Hadron Collider.
No excess above the Standard Model background expectation was observed in 3.2~\ifb\ of analyzed data. 
Results are interpreted within simplified models that assume $R$-parity is conserved and the neutralino is the lightest supersymmetric particle. 
An exclusion limit at the 95\% confidence level on the mass of the
gluino is set at 1.51~\TeV\ for a simplified model incorporating only a gluino octet and the lightest neutralino, assuming the lightest neutralino is massless. For a simplified model involving the strong production of mass-degenerate first- and second-generation squarks, squark masses below 1.03~\TeV\ are excluded for a massless lightest neutralino.
These limits substantially extend the region of supersymmetric parameter space excluded by previous measurements with the ATLAS detector.
}

\AtlasCoverSupportingNote{0L 2-6 jets + MET internal documentation }{https://cds.cern.ch/record/2128394}

\AtlasCoverCommentsDeadline{  }

\AtlasCoverAnalysisTeam{B.~Abeloos, S.~Adachi, G.~Besjes, D.~Bullock, G.~Conti, O.~Dale, I.~Deigaard, L.~Duflot~(*), G.~Fletcher, L.~Heelan, S.~Henrot-Versille, M.~Hodgkinson, P.~Jackson, L.~Lee Jr, N.~Makovec, J.~Mamuzic, Y.~Minami, H.~Moss, Y.~Nakahama, B.~Petersen, A.~Petridis, C.~Rogan, M.~Ronzani, Z.~Rurikova, R.~Woods~Smith, A.~Strubig, D.~Tovey,  M.~Vranjes~Milosavljevic~(*), T.~Yamanaka~(*)}

\AtlasCoverEdBoardMember{Pascal~Pralavorio~(chair), Maximilian~J~Swiatlowski, Benjamin~Henry~Hooberman}

\AtlasCoverEgroupEditors{atlas-SUSY-2015-06-editors@cern.ch}

\AtlasCoverEgroupEdBoard{atlas-SUSY-2015-06-editorial-board@cern.ch}

\begin{document}

\maketitle

\tableofcontents

\clearpage

\section{Introduction}
\label{sec:intro}
Supersymmetry (SUSY)
\cite{Golfand:1971iw,Volkov:1973ix,Wess:1974tw,Wess:1974jb,Ferrara:1974pu,Salam:1974ig}  
is a generalization of space-time symmetries that 
predicts new bosonic partners for the fermions and new fermionic partners for the bosons
of the Standard Model (SM). If $R$-parity is conserved~\cite{Farrar:1978xj}, 
SUSY particles (called sparticles) are produced in pairs and the lightest supersymmetric particle (LSP) is stable and represents a possible dark-matter candidate. 
The scalar partners of the left- and right-handed quarks, the squarks $\squarkL$ and $\squarkR$, mix to form two mass eigenstates  $\tilde{q}_1$ and $\tilde{q}_2$ ordered by increasing mass. Superpartners of the charged and neutral electroweak and Higgs bosons also mix to produce charginos ($\chinopm$) and neutralinos  ($\nino$). Squarks and the fermionic partners of the gluons, the gluinos ($\gluino$), could be produced in strong-interaction processes at the Large Hadron Collider (LHC)   \cite{LHC:2008}  and decay via cascades ending with the stable LSP, which escapes the detector unseen, producing substantial missing transverse momentum ($\bm{E}\mathrm{^{miss}_T}$).

The production of gluinos and squarks is the primary target for early supersymmetry searches in proton--proton ($pp$) collisions at a centre-of-mass energy of 13~\TeV\ at the LHC because of the large expected cross-sections predicted for the production of supersymmetric particles which participate to the strong interaction. 
This document presents a search for these particles in final states containing only hadronic jets and large missing transverse momentum.  
Interest in this final state is motivated by the large number of $R$-parity-conserving models~\cite{Fayet:1976et,Fayet:1977yc} in which squarks (including anti-squarks) and gluinos can be produced in pairs ($\gluino\gluino$, 
$\squark \squark$, $\squark
\gluino$) and can decay through
$\squark \to q
\ninoone$
and $\gluino
\to q \bar{q} \ninoone$ to the lightest neutralino, $\ninoone$, assumed to be the LSP. 
Additional decay modes can include the production of charginos via $\squark\to q\chinopm$ (where $\squark$ and $q$ are of different flavour) and $\gluino\to q\bar{q}\chinopm$. Subsequent chargino decay to $W^{\pm}\ninoone$ can lead to still larger multiplicities of jets. 
The analysis presented here adopts the same analysis strategy as the previous ATLAS search designed for the analysis of the 7~\TeV\ and 8~\TeV\ data collected during Run~1 of the LHC, described in Refs.~\cite{daCosta:2011qk,Aad:2011ib,Aad:2012fqa,0-leptonPaper,summaryPaper}. %
The CMS Collaboration has set limits on similar models in Refs.~\cite{Chatrchyan:2012jx,Chatrchyan:2012lia,Chatrchyan:2013lya,Chatrchyan:2012uea,Chatrchyan:2014lfa,Khachatryan:2015vra}.

In this search, events with reconstructed electrons or muons are rejected to reduce the background from events with neutrinos ($W \rightarrow e\nu,\mu\nu$) and to avoid any overlap with a complementary ATLAS search in final states with one lepton, jets and missing transverse momentum  \cite{1leptonPaper}. The selection criteria are optimized in the $(m_{\gluino}, m_{\ninoone})$ and $(m_{\squark}, m_{\ninoone})$ planes, (where $m_{\gluino}$, $m_{\squark}$ and $m_{\ninoone}$ are the gluino, squark and the LSP masses, respectively) for simplified models  \cite{Alwall:2008ve,Alwall:2008ag,Alves:2011wf} in which all other supersymmetric particles are assigned masses beyond the reach of the LHC.
Although interpreted in terms of SUSY models, the results of this analysis could also constrain any model  
of new physics that predicts the production of jets in association with missing transverse momentum.  

\section{The ATLAS detector and data samples}
\label{sec:detector}
\interfootnotelinepenalty=10000
The ATLAS detector~\cite{Aad:2008zzm} is a multi-purpose detector with a forward-backward symmetric cylindrical
geometry and nearly 4$\pi$ coverage in solid angle.\footnote{
ATLAS uses a right-handed coordinate system with its origin at the nominal
interaction point in the centre of the detector. The positive $x$-axis is defined by the direction from the interaction point to the centre
of the LHC ring, with the positive $y$-axis pointing upwards, while the beam direction defines the $z$-axis. Cylindrical coordinates $(r,\phi)$ are used in the transverse
plane, $\phi$ being the azimuthal angle around the $z$-axis. The pseudorapidity $\eta$ is
defined in terms of the polar angle $\theta$ by $\eta=-\ln\tan(\theta/2)$ and the rapidity is defined as $y = (1/2)\ln[(E+p_z)/(E-p_z)]$ where $E$ is the energy and $p_{\rm z}$ the longitudinal
momentum of the object of interest. The transverse momentum $\pt$, the transverse energy $\ET$ and the missing transverse momentum $\met$ are defined in the $x$--$y$ plane unless stated otherwise. 
}   
The inner tracking detector (ID) consists of pixel and silicon microstrip detectors 
covering the pseudorapidity region $|\eta|<2.5$, surrounded by a transition radiation tracker  
which improves electron identification over the region $|\eta|<2.0$.  The innermost
pixel layer, the insertable B-layer \cite{B-layerRef}, was added between Run~1 and Run~2 of the LHC, at a radius of 33 mm around a new, narrower and thinner, beam pipe. 
The ID is surrounded by a thin superconducting solenoid providing an axial 2~T magnetic field and by
a fine-granularity lead/liquid-argon (LAr) electromagnetic calorimeter covering $|\eta|<3.2$.
A steel/scintillator-tile calorimeter provides hadronic coverage in
the central pseudorapidity range ($|\eta|<1.7$). 
The endcap and forward regions ($1.5<|\eta|<4.9$) of the hadronic calorimeter are made of LAr active layers with either copper or tungsten as the absorber material. 
The muon spectrometer with an air-core toroid magnet system surrounds the calorimeters.
Three layers of high-precision tracking chambers
provide coverage in the range $|\eta|<2.7$, while dedicated chambers allow triggering in the region $|\eta|<2.4$.

The ATLAS trigger system \cite{atlastrigger} consists of two levels; the first level is a hardware-based system, while the second is a software-based system called the High-Level Trigger. The events used in this search were selected using a trigger logic that accepts events with a missing transverse momentum above 70 GeV, calculated using a sum over calorimeter cells. The trigger is 100\% efficient for the event selections considered in this analysis. Auxiliary data samples used to estimate the yields of background events were selected using triggers requiring at least one isolated electron ($\pt>24\GeV$), muon ($\pt>20\GeV$) or photon ($\pt>120\GeV$). To increase the  efficiency at high momenta, additional single-electron and single-muon triggers that do not require any isolation were included with thresholds of $\pt = 60 \GeV$ and $\pt = 50 \GeV$, respectively.

The dataset used in this analysis was collected in 2015 with the LHC colliding proton beams at a centre-of-mass energy of 13~\TeV, with 25 ns bunch spacing. 
The peak delivered instantaneous luminosity was $L = 5.2 \times 10^{33} ~{\rm cm^{-2} s^{-1}}$ and the mean number of additional $pp$ interactions per bunch crossing in the dataset was $\mean{\mu}$ = 14. 
Application of beam, detector and data-quality criteria resulted in a total integrated luminosity of 3.2~\ifb.  The uncertainty in the integrated luminosity is $\pm$5\%. It is derived, following a methodology similar to that detailed in Ref.~\cite{Aad:2013ucp}, from a preliminary calibration of the luminosity scale using a pair of $x$--$y$ beam-separation scans performed in August 2015. 

\section{Monte Carlo simulated samples}
\label{sec:montecarlo}

Simulated Monte Carlo (MC) data samples are used to optimize the selections, estimate backgrounds and assess the sensitivity to specific SUSY signal models.  

SUSY signals are described in this paper by simplified models. They are defined by an effective Lagrangian describing the interactions of a small number of new particles, typically assuming one production process and one decay channel with a 100\% branching fraction. 
Signal samples used to describe squark- and gluino-pair production, followed by the direct\footnote{Direct decays are those where the considered SUSY particles decay directly into SM particles and the LSP.} decays of squarks ($\squark \rightarrow q\ninoone$) and direct ($\gluino \rightarrow q\bar{q} \ninoone$) or one-step\footnote{One-step decays refer to the cases where the decays occur via one  intermediate on-shell SUSY particle.} ($\gluino \rightarrow q\bar{q}'W \ninoone$) decays of gluinos as shown in Figure \ref{fig:feynman_directgrids}, are generated with up to two extra partons in the matrix element using MG5\_aMC@NLO event generator~\cite{Alwall:2014hca} 
interfaced to  \pythia~8.186~\cite{Sjostrand:2014zea}. The CKKW-L merging scheme~\cite{Lonnblad:2011xx} is applied with a scale parameter that is set to a quarter of the mass of the gluino for $\gluino\gluino$ production or of the squark for $\squark\squark$ production. The A14~\cite{A14tune} set of tuned parameters (tune) is used for underlying event together with the NNPDF2.3LO~\cite{Ball:2012cx} parton distribution function (PDF) set. 
The {\sc EvtGen}~v1.2.0 program~\cite{evtgen} is used to describe the properties of the $b$- and $c$- hadron decays in the signal samples and the background samples except those produced with \sherpa~\cite{sherpa2}. 
The signal cross-sections are calculated at next-to-leading order (NLO) in the strong coupling constant, adding the resummation of soft gluon emission at next-to-leading-logarithmic accuracy (NLO+NLL)~\cite{Beenakker:1996ch,Kulesza:2008jb,Kulesza:2009kq,Beenakker:2009ha,Beenakker:2011fu}. The nominal cross-section is taken from an envelope of cross-section predictions using different PDF sets and factorization and renormalization scales, as described in Ref.~\cite{Kramer:2012bx}, considering only light-flavour quarks ($u$, $d$, $s$, $c$). 
Cross-sections are evaluated assuming masses of 450 \TeV\ for the light-flavour squarks in case of gluino- or gluinos in case of squark-pair production. 
The free parameters are $m_{\ninoone}$ and $m_{\squark}$ ($m_{\gluino}$) for gluino-pair (squark-pair) production models.

\begin{figure}[h]
\begin{center}
\subfigure[]{\includegraphics[width=0.25\textwidth]{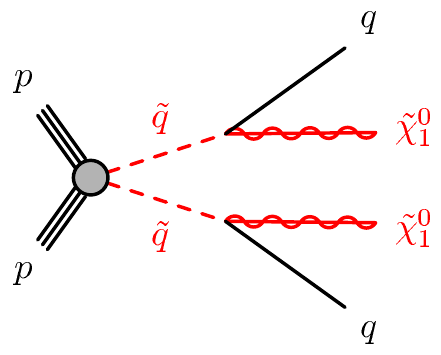}}\hspace{0.05\textwidth}
\subfigure[]{\includegraphics[width=0.25\textwidth]{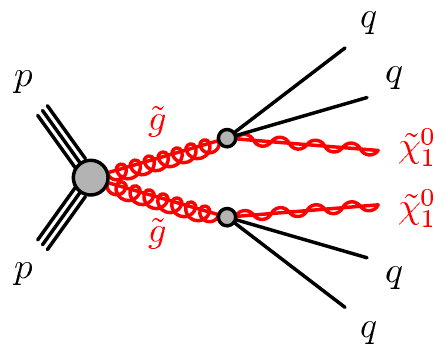}}\hspace{0.05\textwidth}
\subfigure[]{\includegraphics[width=0.25\textwidth]{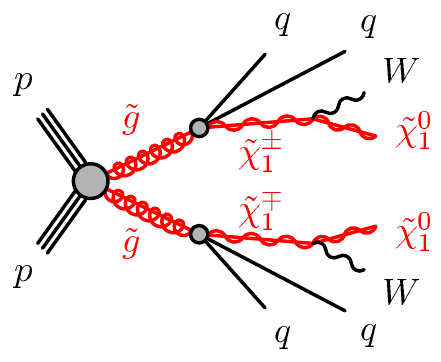}}
\caption{The decay topologies of (a) squark-pair production and (b, c) gluino-pair production, in the simplified models with direct decays of squarks and direct or one-step decays of gluinos. }
\label{fig:feynman_directgrids}
\end{center}                  
\end{figure}

A summary of the SM background processes together with the MC generators, cross-section calculation orders in $\alpha_{\rm s}$, PDFs, parton shower and tunes used is given in Table~\ref{tab:montecarlo}.

\begin{table}[H]
\scriptsize
\caption{The Standard Model background Monte Carlo simulation samples used in this paper. The generators, the order in $\alpha_{\rm s}$ of cross-section calculations used for yield normalization, PDF sets, parton showers and tunes used for the underlying event are shown. }
\begin{center}
\begin{tabular}{| l l c c c c |}
\hline
Physics process & Generator& Cross-section & PDF set & Parton shower & Tune \\
&& normalization & & & \\
\hline
$W(\rightarrow \ell\nu)$ + jets              & \sherpa~2.1.1        & NNLO  &  CT10   &  \sherpa\     & \sherpa~default \\
$Z/\gamma^{*}(\rightarrow \ell \bar \ell)$ + jets & \sherpa~2.1.1         & NNLO  &  CT10   & \sherpa\      & \sherpa~default\\
$\gamma $ + jets & \sherpa~2.1.1         & LO  &    CT10  & \sherpa\   & \sherpa~default\\

$t\bar{t}$              & {\sc Powheg-Box}~v2   & NNLO+NNLL                   &  CT10 &  \pythia~6.428  &\sc{Perugia2012} \\

Single top ($Wt$-channel) & {\sc Powheg-Box} v2  &  NNLO+NNLL  &  CT10 &  \pythia~6.428   & \sc{Perugia2012}\\ 
Single top ($s$-channel)           & {\sc Powheg-Box} v2  & NLO  &  CT10 &  \pythia~6.428   & \sc{Perugia2012}\\
Single top ($t$-channel)           & {\sc Powheg-Box} v1  & NLO  &  CT10f4 &  \pythia~6.428   & \sc{Perugia2012}\\

$t\bar{t}+W/Z/WW$       &  MG5\_aMC@NLO  & NLO  & NNPDF2.3LO & \pythia~8.186 & A14    \\

$WW$, $WZ$, $ZZ$    &  \sherpa~2.1.1       & NLO  &  CT10 & \sherpa\   & \sherpa~default \\
Multi-jet    &  \pythia~8.186       & LO  & NNPDF2.3LO & \pythia~8.186   & A14\\

\hline
\end{tabular}

\label{tab:montecarlo}
\end{center}
\end{table}

The production of $\gamma$, $W$ or $Z$ bosons in association with jets~\cite{ATL-PHYS-PUB-2016-003} is simulated using the \sherpa~2.1.1 generator. For $W$ or $Z$ bosons, the matrix elements are calculated for up to two partons at NLO and up to two additional partons at leading order (LO) using the \textsc{Comix} \cite{comix} and \textsc{OpenLoops} \cite{openloops} matrix-element generators, and merged with the \sherpa\ parton shower \cite{sherpashower} using the ME+PS@NLO prescription \cite{mepsnlo}. 
Events containing a photon in association with jets are generated requiring a photon transverse momentum above 35~\GeV. For these events, matrix elements are calculated at LO with up to three or four partons depending on the $\pt$ of the photon, and merged with the \sherpa\ parton shower using the ME+PS@LO prescription~\cite{Hoeche:2009rj}. 
In both cases ($W/Z$+jets or $\gamma$+jets production), the CT10 PDF set \cite{CT10pdf} is used in conjunction with dedicated parton shower-tuning developed by the authors of \sherpa. The $W/Z$ + jets events are normalized to their NNLO cross-sections \cite{Catani:2009sm}. For the $\gamma$+jets process the LO cross-section, taken directly from the \sherpa\ MC generator, is multiplied by a correction factor as described in Section~\ref{sec:background}. 

For the generation of $t\bar{t}$ and single-top processes in the $Wt$ and $s$-channel~\cite{ATL-PHYS-PUB-2016-004} the \textsc{Powheg-Box} v2 \cite{powheg-box} generator is used with the CT10 PDF set. The electroweak (EW) $t$-channel single-top events are generated using the \textsc{Powheg-Box} v1 generator. This generator uses the four-flavour scheme for the NLO matrix-element calculations together with the fixed four-flavour PDF set CT10f4~\cite{CT10pdf}. For this process, the decay of the top quark is simulated using {\sc MadSpin} tool \cite{10a} preserving all spin correlations, while for all processes the parton shower, fragmentation, and the underlying event are generated using \pythia~6.428 \cite{pythia6} with the CTEQ6L1 \cite{Pumplin:2002vw} PDF set and the corresponding {\sc Perugia 2012} tune (P2012) \cite{perugia}. The top quark mass is set to 172.5~\GeV. 
The $h_{\rm damp}$ parameter, which controls the $\pt$ of the first additional emission beyond the Born configuration, is set to the mass of the top quark. The main effect of this is to regulate the high-$\pt$ emission against which the ttbar system recoils \cite{ATL-PHYS-PUB-2016-004}.
The $t\bar{t}$ events are normalized to the NNLO+NNLL ~\cite{Czakon:2013goa,Czakon:2011xx}. The $s$- and $t$-channel single-top events are normalized to the NLO cross-sections \cite{Aliev:2010zk,Kant:2014oha}, and the $Wt$-channel  single-top events are normalized to the NNLO+NNLL~\cite{Kidonakis:2010ux,Kidonakis:2011wy}. 

For the generation of $t\bar{t}$ + EW processes ($t\bar{t} + W/Z/WW$)~\cite{ATL-PHYS-PUB-2016-005}, the MG5\_aMC@NLO~\cite{Alwall:2014hca} generator at LO interfaced to the \pythia~8.186 parton-shower model is used, with up to two ($t\bar{t}+W$), one ($t\bar{t}+Z$) or no ($t\bar{t}+WW$) extra partons included in the matrix element. The ATLAS underlying-event tune A14 is used together with the NNPDF2.3LO PDF set.  The events are normalized to their respective NLO cross-sections~\cite{Lazopoulos:2008de,Campbell:2012dh}. 

Diboson processes ($WW$, $WZ$, $ZZ$)~\cite{ATL-PHYS-PUB-2016-002} are simulated using the  \sherpa~2.1.1 generator. For processes with four charged leptons (4$\ell$), three charged leptons and a neutrino (3$\ell$+1$\nu$) or two charged leptons and two neutrinos (2$\ell$+2$\nu$), the matrix elements contain all diagrams with four electroweak vertices, and are calculated for up to one (4$\ell$, 2$\ell$+2$\nu$) or no partons (3$\ell$+1$\nu$) at NLO and up to three partons at LO using the \textsc{Comix} and \textsc{OpenLoops} matrix-element generators, and merged with the \sherpa\ parton shower using the ME+PS@NLO prescription. 
For processes in which one of the bosons decays hadronically and the other leptonically, matrix elements are calculated for up to one ($ZZ$) or no ($WW$, $WZ$) additional partons at NLO and for up to three additional partons at LO using the \textsc{Comix} and \textsc{OpenLoops} matrix-element generators, and merged with the \sherpa\ parton shower using the ME+PS@NLO prescription. In all cases, the CT10 PDF set is used in conjunction with a dedicated parton-shower tuning developed by the authors of  \sherpa. The generator cross-sections are used in this case.

The multi-jet background is generated with \pythia~8.186 using the A14 underlying-event tune and the NNPDF2.3LO parton distribution functions. 

For all Standard Model background samples the response of the detector to particles is modelled with a full ATLAS detector simulation \cite{:2010wqa} based on \textsc{Geant4} \cite{Agostinelli:2002hh}. 
Signal samples are prepared using a fast simulation based on a parameterization of the performance of the ATLAS electromagnetic and hadronic calorimeters \cite{ATLAS:2010bfa} and on \textsc{Geant4} elsewhere.  

All simulated events are overlaid with multiple $pp$ collisions simulated with the soft QCD processes of \pythia~8.186 using the A2 tune  \cite{A14tune} and the MSTW2008LO parton distribution functions \cite{Martin:2009iq}. The simulations are not reweighted to match the distribution of the mean number of interactions observed in data. It was checked that the effect of such pile-up reweighting is completely negligible.

\section{Object reconstruction and identification}
\label{sec:objects}
The reconstructed primary vertex of the event is required to be consistent with the luminous region and to have at least two associated tracks with $\pt > 400$~\MeV. When more than one such vertex is found, the vertex with the largest  $\sum \pt^2$ of the associated tracks is chosen.

Jet candidates are reconstructed using the anti-$k_{t}$ jet clustering algorithm~\cite{Cacciari:2008gp,Cacciari:2005hq} with jet 
radius parameter of $0.4$ and starting from clusters of calorimeter cells \cite{Lampl:2008}. 
The jets are corrected for energy from pile-up %
using the method suggested in Ref.~\cite{Cacciari:2007fd}: a contribution equal to the product of the jet area and the median energy density of the event is subtracted from the jet energy \cite{ATLAS-CONF-2013-083}.
Further corrections, referred to as the jet energy scale corrections, are derived from MC simulation and data and used to calibrate on average the energies of jets to the scale of their constituent particles \cite{JetCalibRun2}. 
Only jet candidates with $\ourpt > 20\GeV$ and $|\eta|<2.8$ after all corrections are retained. 
An algorithm based on boosted decision trees, `MV2c20' \cite{ATL-PHYS-PUB-2015-022}, is used to identify jets containing a $b$-hadron ($b$-jets), with an operating point corresponding to an efficiency of 77\% in simulated $t\bar{t}$ events, along with a rejection factor of 140 for gluon and light-quark jets and of 4.5 for charm jets~\cite{ATL-PHYS-PUB-2015-022,ATL-PHYS-PUB-2015-039}. Candidate $b$-tagged jets are required to have $\ourpt > 50\GeV$ and $|\eta|<2.5$. 
Events with jets originating from detector noise and non-collision background are rejected if the jets fail to satisfy the `LooseBad' quality criteria, or if at least one of the two leading jets with $\pt >100 \GeV$ fails to satisfy the `TightBad' quality criteria, both described in Ref.~\cite{Jets2015}. 
These selections affect less than 1\% of the events used in the search. 

Two different classes of reconstructed lepton candidates (electrons or muons) are used in this analysis. When selecting samples used for the search, events containing a `baseline' electron or muon are rejected. The selections applied to identify baseline leptons are designed to maximize the efficiency with which $W$+jets and top quark background events are rejected. When selecting `control region' samples for the purpose of estimating residual $W$+jets and top quark backgrounds, additional requirements are applied to leptons to ensure greater purity of the these backgrounds. These leptons are referred to as `high-purity' leptons below and form a subset of the baseline leptons.

Baseline muon candidates are formed by combining information from the muon spectrometer and inner tracking detectors as described in Ref.~\cite{MuonPerfRun2} and are required to have $\ourpt > 10 \GeV$ and $|\eta| <2.7$.  High-purity muon candidates must additionally have $|\eta|<2.4$, the significance of the transverse impact parameter with respect to the primary vertex, $|d_0^{\mathrm{PV}}|/\sigma(d_0^{\mathrm{PV}}) <$ 3, the longitudinal impact parameter with respect to the primary vertex  $|z_0^{\mathrm{PV}} \mathrm{sin}(\theta)|<$ 0.5~mm, and to satisfy `GradientLoose' isolation requirements described in Ref.~\cite{MuonPerfRun2} which rely on the use of tracking-based and calorimeter-based variables and implement a set of $\eta$- and $\pt$-dependent criteria.  The leading muon is also required to have $\pt > 25\GeV$. 

Baseline electron candidates are reconstructed from an isolated electromagnetic calorimeter energy deposit matched to an ID track and are required to have $\ourpt > 10\GeV$, $|\eta| < 2.47$, and to satisfy `Loose' likelihood-based identification criteria described in Ref.~\cite{ATL-PHYS-PUB-2015-041}.  
High-purity electron candidates additionally must satisfy `Tight' selection criteria described in Ref.~\cite{ATL-PHYS-PUB-2015-041}, and the leading electron must have $\pt>25 \GeV$. They are also required to have $|d_0^{\mathrm{PV}}|/\sigma(d_0^{\mathrm{PV}}) <$ 5, $|z_0^{\mathrm{PV}} \mathrm{sin}(\theta)|<$ 0.5~mm, and to satisfy similar isolation requirements as those applied to high-purity muons.

After the selections described above, ambiguities between candidate jets with $|\eta|<2.8$ and leptons are resolved as follows: first, any such jet candidate lying within a distance $\Delta
R\equiv\sqrt{(\Delta y)^2+(\Delta\phi)^2}=0.2$ of a baseline electron is discarded;
then  any baseline lepton candidate remaining within a distance
$\Delta R =0.4$ of any surviving jet candidate is discarded, except in the case where the lepton is a muon (which can radiate a photon and be misidentified as a jet) and the number of tracks associated with the jet is less than three. 

Additional ambiguities between electrons and muons in a jet, originating from the decays of hadrons, are resolved to avoid double counting and/or remove non-isolated leptons: the electron is discarded if a baseline electron and a baseline muon share the same ID track. If two baseline electrons are within $\Delta R$ = 0.05, the electron with the lowest $\pt$ is discarded. 

The measurement of the missing transverse momentum vector
$\bm{E}\mathrm{^{miss}_T}$ (and its magnitude $\ourmagptmiss$) is based on
the calibrated transverse momenta of all electron, muon, photon and jet candidates and 
all tracks originating from the primary vertex and not associated with such objects~\cite{MET2015}.

Reconstructed photons, although not used in the main signal-event selection, are selected in the region  used to constrain the $Z$+jets background, as explained in Section~\ref{sec:background}. Photon candidates are required to satisfy $\ourpt > 130\GeV$ and $|\eta| < 2.37$, to satisfy photon shower shape and electron rejection criteria \cite{ATLAS-CONF-2012-123}, and to be isolated.
Ambiguities between candidate jets and photons (when used in the event selection) are resolved by discarding any jet candidates lying within $\Delta R$ = 0.4 of a photon candidate. %
Additional selections to remove ambiguities between electrons or muons and photons are applied such that the photon is discarded if it is within $\Delta R$ = 0.4 of an electron or muon.

Corrections derived from data control samples are applied to account for differences between data and simulation for the lepton trigger and reconstruction efficiencies, the lepton momentum/energy scale and resolution, and for the efficiency and mis-tag rate of the $b$-tagging algorithm.

\section{Analysis strategy and fit description}
\label{sec:strategy}

To search for a possible signal, selections are defined to enhance the signal relative to the SM background. These signal region (SR) selections are optimized to maximize the expected significance for each model considered using MC simulation for the signal and the SM backgrounds. 
To estimate the SM backgrounds in a consistent and robust fashion, corresponding control regions (CRs) are defined for each of the signal regions. 
They are chosen to be non-overlapping with the SR selections in order to provide independent data samples enriched 
in particular background sources, and are used to normalize the background MC simulation. The CR selections are optimized to have negligible SUSY signal contamination for the models near the previously excluded boundary \cite{0-leptonPaper}, while minimizing the systematic uncertainties arising from the extrapolation of the CR event yields to estimate backgrounds in the SR. 
Cross-checks of the background estimates are performed with data in several validation regions (VRs) selected with requirements such that these regions do not overlap with the CR and SR selections, again with a low expected signal contamination. 

To extract the final results, three different classes of likelihood fit are employed: background-only, model-independent and model-dependent fits  \cite{HFpaper}. 
A background-only fit is used to estimate the background yields in each SR. The fit is performed using as constraints only the observed event yields from the CRs associated with the SR, but not the SR itself. It is assumed that signal events from physics beyond the Standard Model (BSM) do not contribute to these yields. 
The scale factors ($\mu_{W\rm+jets}$, $\mu_{Z\rm+jets}$, $\mu_{\rm Top}$, $\mu_{\rm Multi-jet}$) are fitted in each CR attached to a SR. 
The expected background in the SR is based on the yields predicted by simulation, corrected by the scale factors derived from the fit. %
The systematic uncertainties and the MC statistical uncertainties in the expected values are included in the fit as nuisance parameters which are constrained by Gaussian distributions with widths corresponding to the sizes of the uncertainties considered and by Poisson distributions, respectively.  
The background-only fit is also used to estimate the background event yields in the VRs.

If no excess is observed, a model-independent fit is used to set upper limits on the number of BSM signal events in each SR. This fit proceeds in the same way as the background-only fit, except that the number of events observed in the SR is added as an input to the fit, and the BSM signal strength, constrained to be non-negative, is added as a free parameter. The observed and expected upper limits at 95\% confidence level (CL) on the number of events from BSM phenomena for each signal region ($S_{\rm obs}^{95}$ and $S_{\rm exp}^{95}$) are derived using the $CL_{\rm s}$ prescription \cite{Read:2002hq}, neglecting any possible signal contamination in the control regions. These limits, when normalized by the integrated luminosity of the data sample, may be interpreted as upper limits on the visible cross-section of BSM physics ($\langle\epsilon\sigma\rangle_{\rm obs}^{95}$), where the visible cross-section is defined as the product of production cross-section, acceptance and efficiency. The model-independent fit is also used to compute the one-sided $p$-value ($p_0$) of the background-only hypothesis, which quantifies the statistical significance of an excess.

Finally, model-dependent fits are used to set exclusion limits on the signal cross-sections for specific SUSY models. Such a fit proceeds in the same way as the model-independent fit, except that both the yield in the signal region and the signal contamination in the CRs are taken into account. Correlations between signal and background systematic uncertainties are taken into account where appropriate. 
Signal-yield systematic uncertainties due to detector effects and the theoretical uncertainties in the signal acceptance are included in the fit.

\section{Event selection and signal regions definitions}
\label{sec:selection}
Due to the high mass scale expected for the SUSY models considered in this study, the `effective mass', $\meff$, is a powerful discriminant between the signal and most SM backgrounds. 
When selecting events with at least $N_{\rm j}$ jets, $\meff(N_{\rm j})$ is defined to be the scalar sum of the transverse momenta of the leading $N_{\rm j}$ jets and \met{}. %
Requirements placed on $\meff$ and \met{} form the basis of this search by strongly suppressing the multi-jet background where jet energy mismeasurement generates missing transverse momentum. The final signal selection uses requirements on both $\meff({\rm incl.})$, which sums over all jets with $\ourpt>50 \GeV$ and \met{}, which is required to be larger than 200 \GeV.

Following the object reconstruction described in Section~\ref{sec:objects}, events are discarded if a baseline electron or muon with $\pt>10\GeV$ remains, or if they contain a jet failing to satisfy quality selection criteria designed to suppress detector noise and non-collision backgrounds (described in Section~\ref{sec:objects}). 
Events are also rejected if no jets with $\pt >50\GeV$ are found. %
Reconstructed photons and hadronically decaying $\tau$-leptons are not used in SR selections. 

In order to maximize the sensitivity in the $(m_{\gluino},m_{\squark})$ plane, a variety of signal regions are defined. Squarks typically generate at least one jet in their decays, for instance through $\squark \to q \ninoone$, while gluinos typically generate at least two jets, for instance through $\gluino\to q \bar{q} \ninoone$. Processes contributing to $\squark\squark$ and $\gluino\gluino$ final states therefore lead to events containing at least two or four jets, respectively. Decays of heavy SUSY and SM particles produced in longer $\squark$ and $\gluino$ decay cascades (e.g. $\chinoonepm\to qq'\ninoone$) tend to further increase the jet multiplicity in the final state. 

Seven inclusive SRs characterized by increasing minimum jet multiplicity from two to six, are defined in Table~\ref{tab:srdefs}.  Some of them require the same jet-multiplicity, but are distinguished by increasing background rejection, ranging from `loose' (labelled `l') to `tight' (labelled `t'). 

In each region, different thresholds are applied on jet momenta and on $\ourdeltaphifull$, which is defined to be the smallest azimuthal separation between $\bm{E}\mathrm{^{miss}_T}$ and the momenta of any of the reconstructed jets with $\ourpt>50 \GeV$. Requirements on $\ourdeltaphifull$ and $\ourmagptmiss/\meff(N_{\rm j})$ are designed to reduce the background from multi-jet processes.  
For the SRs which are optimized for squark-pair (gluino-pair) production followed by the direct decay of squarks (gluinos), the selection requires $\ourdeltaphifull>0.8$ ($\ourdeltaphifull>0.4$) using up to three leading jets present in the event. For the SRs requiring at least four jets in the final state, an additional requirement $\ourdeltaphifull>0.2$ is placed on all jets. 
Signal region 2jm makes use of the presence of jets due to initial-state radiation by requiring a higher $\pt$ threshold for the most energetic jet in the event, and is optimized to target models with small mass differences between the SUSY particles (compressed scenarios).  

In the 2-jet SRs the requirement on $\met/\meff(N_{\rm j})$ is replaced by a requirement on $\MET/\sqrt{H_{\rm T}}$ (where $H_{\rm T}$ is defined as the scalar sum of the transverse momenta of all jets), which was found to lead to enhanced sensitivity to models characterized by $\squark\squark$ production. In the other regions, additional suppression of background processes is based on the aplanarity variable, which is defined as $A = 3/2 \lambda_3$, where $\lambda_3$ is the smallest eigenvalue of the normalized momentum tensor of the jets~\cite{Chen:2011aa}.

\begin{table}[H]
\caption{\label{tab:srdefs} Selection criteria and targeted signal model used to define each of the signal regions in the analysis. Each SR is labelled with the inclusive jet multiplicity considered (`2j', `4j' etc.) together with the degree of background rejection. The latter is denoted by labels `l' (`loose'), `m' (`medium') and `t' (`tight'). The $\met/\meff(N_{\rm j})$ cut in any $N_{\rm j}$-jet channel uses a value of $\meff$ constructed from only the leading $N_{\rm j}$ jets ($\meff(N_{\rm j})$).  However, the final $\meff({\rm incl.})$ selection, which is used to define the signal regions, includes all jets with $\pt>50\GeV$. }
\footnotesize
\begin{center}\renewcommand\arraystretch{1.4}
\begin{tabular}{|l|c |c|c| c|c| c| c|}
\hline
\multirow{2}{*}{Requirement}      &\multicolumn{7}{c|}{Signal Region} \\
\cline{2-8}
  & {\bf 2jl} & {\bf 2jm} & {\bf 2jt} & {\bf 4jt} & {\bf 5j} & {\bf 6jm} & {\bf 6jt}\\
 \hline
Targeted signal     & \multicolumn{3}{c|}{$\squark\squark$, $\squark \rightarrow q \ninoone$}  
                                                       & \multicolumn{2}{c|}{$\gluino\gluino$, $\gluino \rightarrow qq \ninoone$} 
                                                       & \multicolumn{2}{c|}{$\gluino\gluino$, $\gluino \rightarrow q\bar{q}'W \ninoone$} \\
\hline
\met [\GeV] $>$&\multicolumn{7}{c|}{ 200 }\\ \hline
$\pt(j_1)$ [\GeV] $>$& 200 & 300 &   \multicolumn{5}{c|}{ 200 } \\ \hline
$\pt(j_2)$ [\GeV] $>$& 200 & 50 & 200 &  \multicolumn{4}{c|}{ 100 }\\ \hline
$\pt(j_3)$ [\GeV] $>$&\multicolumn{3}{c|}{--} &  \multicolumn{4}{c|}{ 100 } \\ \hline 
$\pt(j_4)$ [\GeV] $>$&\multicolumn{3}{c|}{--} &  \multicolumn{4}{c|}{ 100 }  \\ \hline 
$\pt(j_5)$ [\GeV] $>$&\multicolumn{4}{c|}{--} &  \multicolumn{3}{c|}{ 50 } \\ \hline 
$\pt(j_6)$ [\GeV] $>$&\multicolumn{5}{c|}{--} &  \multicolumn{2}{c|}{ 50 }  \\ \hline 
$\ourdeltaphishort(\textrm{jet}_{1,2,(3)}, \bm{E}\mathrm{^{miss}_T})_\mathrm{min}$ $>$ & 0.8 & 0.4 & 0.8 &\multicolumn{4}{c|}{0.4} \\ \hline
$\ourdeltaphishort(\textrm{jet}_{i>3}, \bm{E}\mathrm{^{miss}_T})_\mathrm{min}$ $>$ &\multicolumn{3}{c|}{--}  &\multicolumn{4}{c|}{0.2} \\ \hline

$\met/\sqrt{H_{\rm T}}$ [\GeV$^{1/2}$] $>$ & \multicolumn{2}{c|}{15} & 20 &\multicolumn{4}{c|}{--}  \\ \hline
Aplanarity $>$ &\multicolumn{3}{c|}{--} & \multicolumn{4}{c|}{0.04}  \\ \hline
$\met/\meff(N_{\rm j})$ $>$ &\multicolumn{3}{c|}{--} & 0.2 & \multicolumn{2}{c|}{0.25} & 0.2 \\ \hline
$ \meff({\rm incl.})$ [\GeV] $>$ & 1200 &1600 & 2000 & 2200 & 1600 & 1600 & 2000 \\ \hline
\end{tabular}
\end{center}
\end{table}

\section{Background estimation and validation}
\label{sec:background}

Standard Model background processes contribute to the event counts in
the signal regions. The dominant sources are: $Z+$jets, $W+$jets, top quark
pairs, single top quarks, dibosons and multi-jet production.
Diboson production is estimated with MC simulated data normalized to NLO cross-section predictions, as described in Section~\ref{sec:montecarlo}. 
Most of the $W$+jets background is composed of $W\to \tau\nu$ events in which the $\tau$-lepton decays to hadrons, with additional contributions from $W\to e\nu, \mu\nu$ events in which no baseline electron or muon is reconstructed.
The largest part of the $Z$+jets background comes from the irreducible
component in which $Z\to\nu\bar\nu$ decays generate large $\ourmagptmiss$. 
Top quark pair production followed by semileptonic decays, in particular $t \bar t \to b \bar b \tau \nu q q' $ (with the $\tau$-lepton decaying to hadrons), as well as single-top-quark events,
can also generate large $\ourmagptmiss$ and satisfy the jet
and lepton-veto requirements. %
The multi-jet background in the signal regions is due to missing transverse momentum from misreconstruction of jet energies in the calorimeters, as well as neutrino production in semileptonic decays of heavy-flavour hadrons. After applying the requirements based on $\ourdeltaphifull$ and $\ourmagptmiss/\meff(N_{\rm j})$ listed in Table~\ref{tab:srdefs} the remaining multi-jet background is negligible. %

\begin{table}[H]
\caption{\label{tab:crdefs} Control regions used in the analysis. Also listed are the main targeted background in the SR in each case, the process used to model the background, and the main CR requirement(s) used to select this process. The transverse momenta of high-purity leptons (photons) used to select CR events must exceed 25 (130)~\GeV. The jet $\pt$ thresholds and $\meff({\rm incl.})$ selections match those used in the corresponding SRs.}
  \footnotesize
  \begin{center}\renewcommand\arraystretch{1.2}
    \begin{tabular}{| l  c  c  c |}
      \hline
      CR & SR background &  CR process & CR selection \\ \hline
CR$\gamma$ & $Z(\to\nu\bar\nu)$+jets & $\gamma$+jets & Isolated photon \\
CRQ & Multi-jet & Multi-jet & SR with reversed requirements on (i) $\ourdeltaphifull$  \\
& & & and (ii) $\met/\meff(N_{\rm j})$ or $\met/\sqrt{H_{\rm T}}$\\
CRW & $W(\to\ell\nu)$+jets & $W(\to\ell\nu)$+jets & 30~\GeV $<m_{\rm T}(\ell,\met) < 100$~\GeV, $b$-veto\\
CRT & $t\bar{t}$(+EW) and single top & $t\bar{t}\to b\bar{b}qq'\ell\nu$ & 30~\GeV $<m_{\rm T}(\ell,\met) < 100$~\GeV, $b$-tag\phantom{o}\\ 
\hline
\end{tabular}
  \end{center}
\end{table}

In order to estimate the backgrounds in a consistent and robust fashion, four control regions are defined for each of the seven signal regions, giving 28 CRs in total. 
The CR selections are optimized to maintain adequate statistical precision while minimizing the systematic uncertainties arising from the extrapolation of the CR event yield to estimate the background in the SR. This latter requirement is addressed through the use of CR jet $\pt$ thresholds and $\meff$(incl.) selections which match those used in the SR.
The CR definitions are listed in Table~\ref{tab:crdefs}. 

The CR$\gamma$ region is used to estimate the contribution of 
$Z(\to\nu\bar\nu)$+jets background events to each SR by selecting a sample of $\gamma$+jets events with $\pt(\gamma)>130\GeV$ and then treating the reconstructed photon as contributing to $\met$. For $\pt(\gamma)$ significantly larger than $m_Z$ the kinematic properties of such events strongly resemble those of $Z$+jets events \cite{Aad:2012fqa}. In order to reduce the theoretical uncertainties associated with the $Z/\gamma^*$+jets background expectations in SRs arising from the use of LO $\gamma$+jets cross-sections, a correction factor is applied to the CR$\gamma$ events. This correction factor, $\kappa =1.5 \pm 0.1$, is determined by comparing CR$\gamma$ observations with those in a highly populated auxiliary control region defined by selecting events with two electrons or muons for which the invariant mass lies within 25~\GeV\ of the mass of the $Z$ boson, satisfying $200\GeV < | \bm{E}\mathrm{^{miss}_T} + \bm{p}\mathrm{_T}(\ell\bar\ell) | < 300 \GeV$, together with at least two jets.

The CRW and CRT regions aim to select samples rich in $W(\to \ell \nu)$+jets and semileptonic $t\bar{t}$ background events respectively. Consequently, they differ in their number of $b$-jets (zero or greater or equal to one respectively) but apply the same selection requirements on the transverse mass $m_{\rm T}$ formed by the $E_{\rm T}^{\mathrm{miss}}$ and a high-purity lepton with $p_{\rm T}$ > 25 \GeV. 
These samples are used to estimate respectively the $W$+jets and combined $t\bar{t}$ and single-top background populations, treating the lepton as a jet with the same momentum to model background events in which a hadronically decaying $\tau$-lepton is produced or events in which no baseline electron or muon is reconstructed because it is outside the detector acceptance or below the required $\pt$ threshold. The CRW and CRT selections omit the SR selection requirements on $\ourdeltaphifull$ or $\met/\meff(N_{\rm j})$ ($\met/\sqrt{H_{\rm T}}$ where appropriate) in order to increase the number of CR data events without significantly increasing the theoretical uncertainties associated with the background estimation procedure. 

The CRQ region uses reversed selection requirements on $\ourdeltaphifull$ and on $\met/\meff(N_{\rm j})$ (or $\met/\sqrt{H_{\rm T}}$ where appropriate) to produce samples enriched in multi-jet background events.

As an example, the $ \meff({\rm incl.})$ distributions in control regions associated with SR 4jt are shown in Figure~\ref{fig:cr4j_Meff}. 
In all CRs, the data are consistent with the pre-fit MC background prediction within uncertainties, although the overall normalization is lower by approximately one standard deviation. %

\begin{figure}
\begin{center}
\subfigure[]{\includegraphics[width=0.45\textwidth]{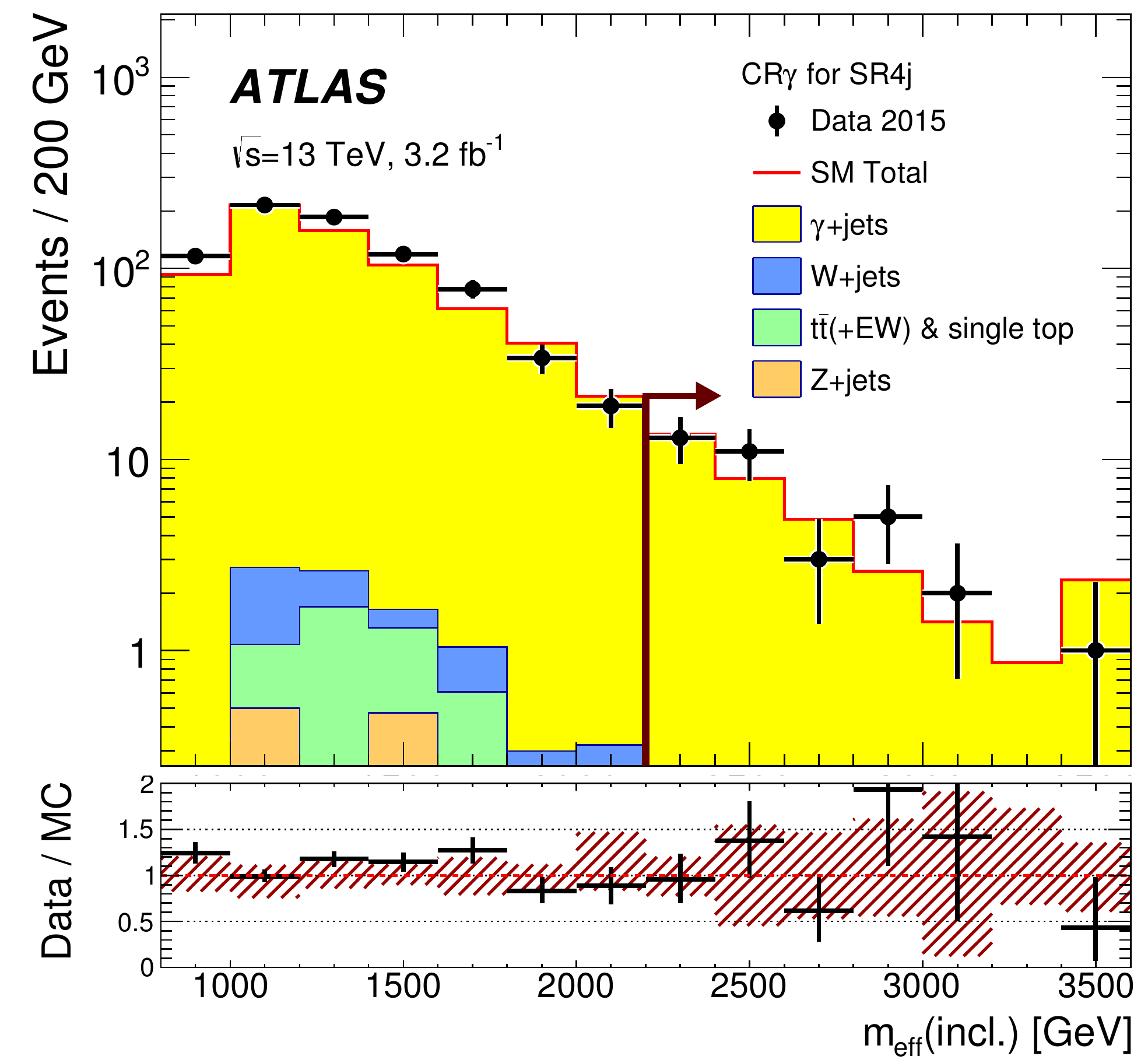}}\\
\subfigure[]{\includegraphics[width=0.45\textwidth]{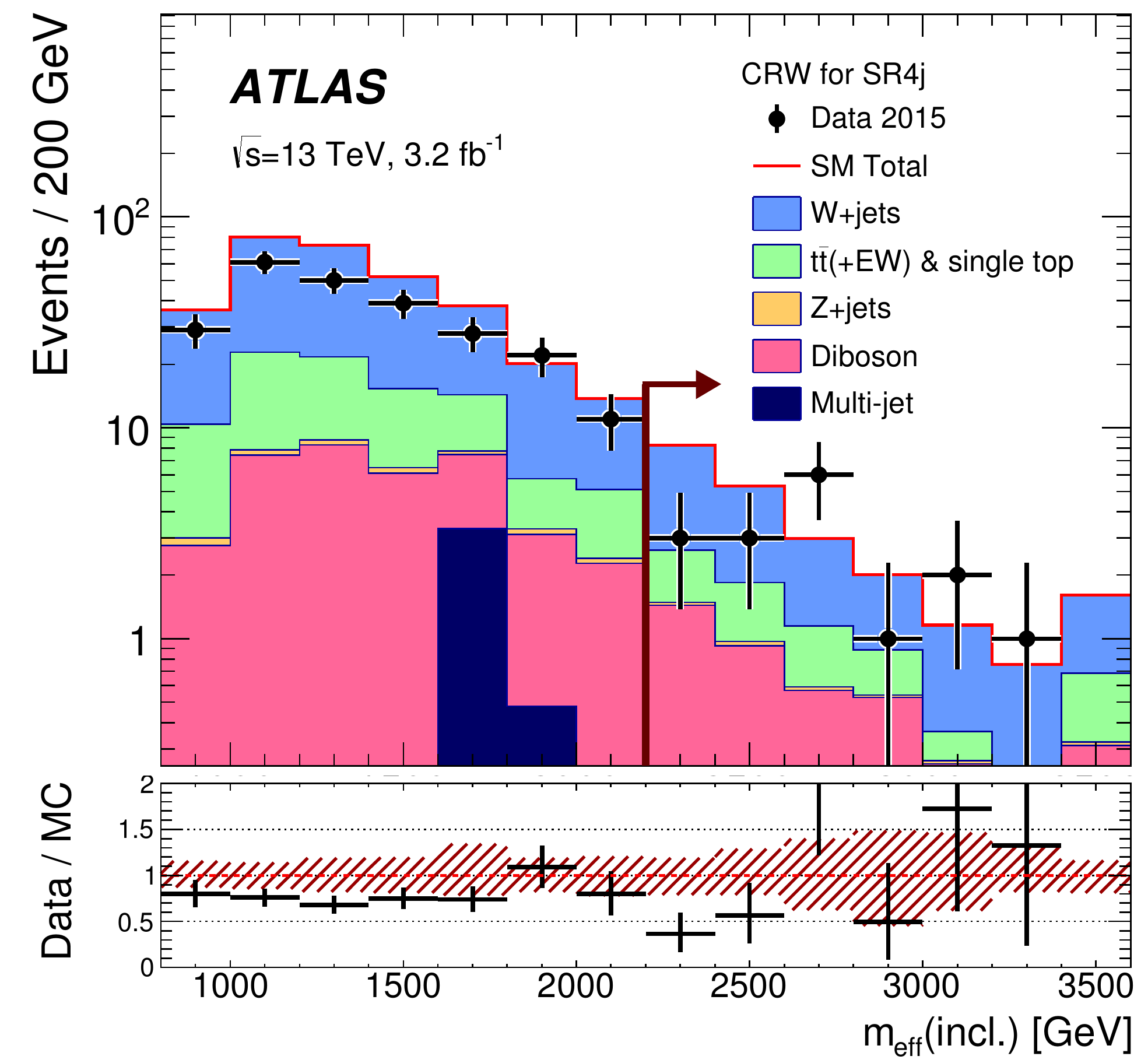}}
\subfigure[]{\includegraphics[width=0.45\textwidth]{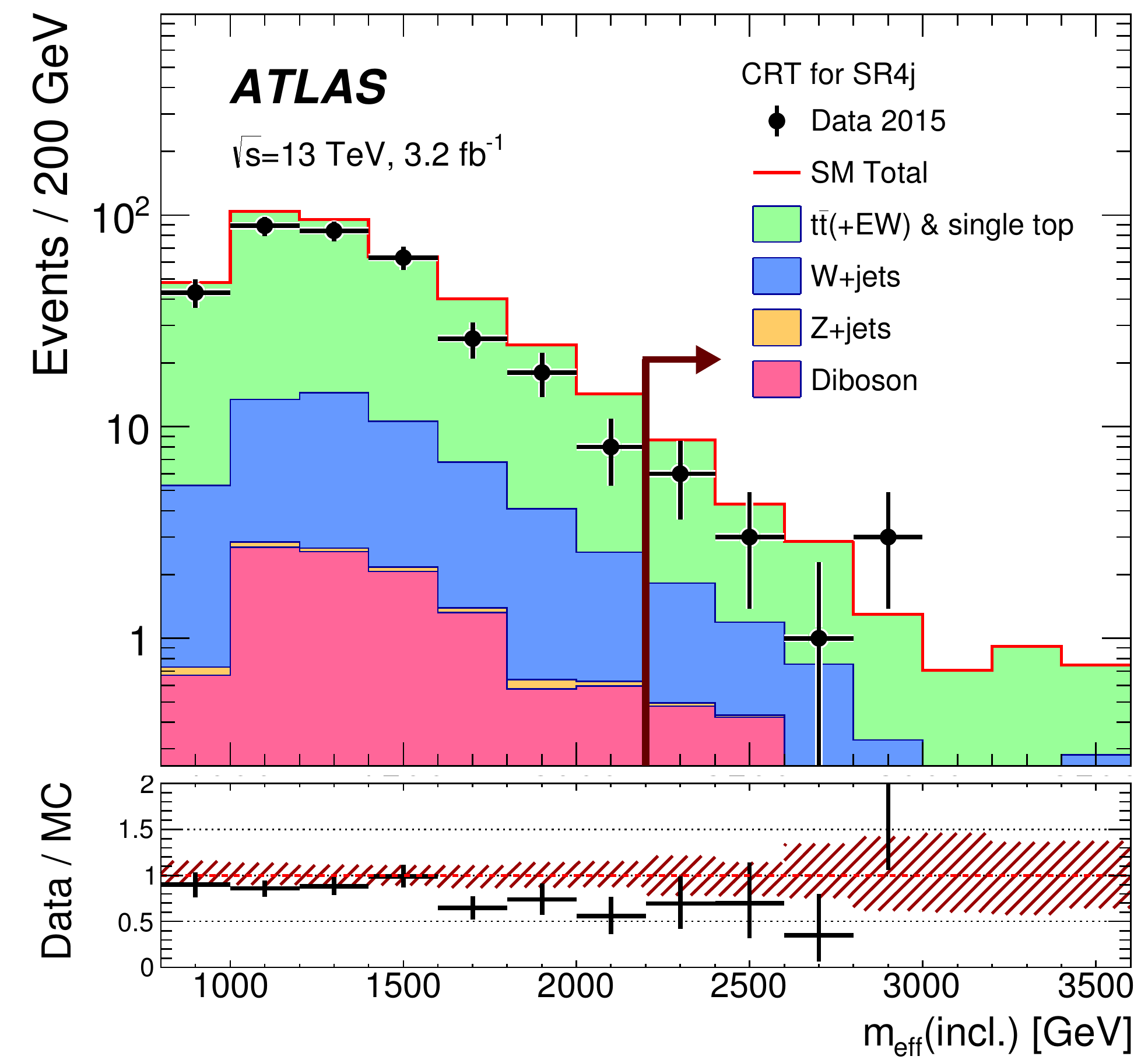}}
\end{center}
\caption{\label{fig:cr4j_Meff}Observed $\meff({\rm incl.})$ distributions in control regions (a) CR$\gamma$, (b) CRW and (c) CRT after selecting events with $\met > 200 \GeV$ and at least four energetic jets with the corresponding transverse momenta as indicated in Table~\ref{tab:srdefs} for SR 4jt. No selection requirements on $\ourdeltaphifull$ or $\met/\meff(N_{\rm j})$ are applied in these distributions. The arrows indicate the values at which the requirements on $\meff({\rm incl.})$ are applied. The histograms denote the pre-fit MC background expectations, normalized to cross-section times integrated luminosity. The last bin includes the overflow. In the lower panels the hatched (red) error bands denote the combined experimental, MC statistical and theoretical modelling uncertainties.   
}
\end{figure}

The background estimation procedure is validated by comparing the numbers of events observed in the VRs to the corresponding SM background expectations obtained from the background-only fits. Several VR samples are selected with requirements distinct from those used in the CRs, which maintain a low probability of signal contamination. 

The CR$\gamma$ estimates of the $Z(\to \nu\bar{\nu})$+jets background are validated using the samples of $Z(\to\ell\bar\ell)$+jets events selected by requiring high-purity lepton pairs of opposite sign and identical flavour for which the dilepton invariant mass lies within 25~\GeV\ of the mass of the $Z$ boson (VRZ). In VRZ, the leptons are treated as contributing to $\met$. 

The CRW and CRT estimates of the $W$+jets and top quark background are validated with the same CRW and CRT selections, but reinstating the requirement on $\ourdeltaphifull$ and $\met/\meff(N_{\rm j})$ (or $\met/\sqrt{H_{\rm T}}$ as appropriate), and treating the lepton either as a jet (VRW, VRT) or as contributing to $\met$ (VRW$\nu$, VRT$\nu$). 

The CRQ estimates of the multi-jet background are validated with VRs for which the CRQ selection is applied, but with the SR $\met/\meff(N_{\rm j})$ ($\met/\sqrt{H_{\rm T}}$) requirement reinstated (VRQa), or with a requirement of an intermediate value of $\ourdeltaphifull$ applied (VRQb). 

The results of the validation procedure are shown in Figure~\ref{fig:VRpulls}. The entries in the matrix are the differences between the numbers of observed and expected events expressed as fractions of the one-standard deviation $(1\sigma)$ uncertainties on the latter. Most VR observations lie within $1\sigma$ of the background expectations, with the largest discrepancy out of 49 VRs being $-1.5\sigma$ the CRQb associated with the SR 4jt.

\begin{figure}[H]
\begin{center}
\includegraphics[width=0.6\textwidth]{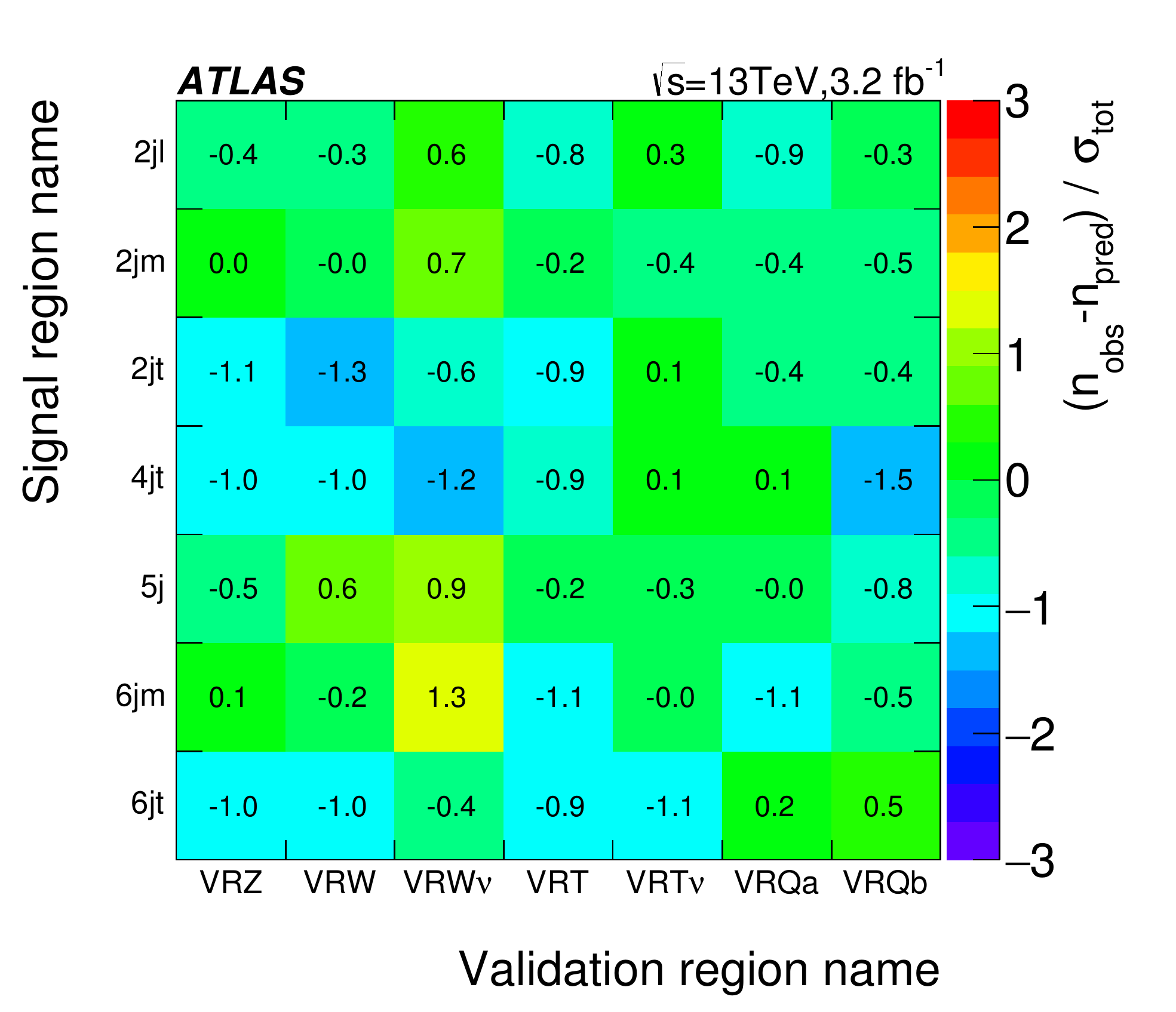}
\end{center}
\caption{\label{fig:VRpulls} Differences between the numbers of observed events in data and the SM background predictions for each VR, 
expressed as a fraction of the total uncertainty which combines the uncertainty on the background expectations, and the expected statistical uncertainty of the test obtained from the number of expected events. }
\end{figure}

\section{Systematic uncertainties}
\label{sec:systematics}
Systematic uncertainties in background estimates arise from the use of extrapolation factors which relate observations in the control regions to background expectations in the signal regions, and from the MC modelling of minor backgrounds.

The overall background uncertainties, detailed in Table~\ref{tab:BreakdownSysSRCompressed}, range from 8\% in SR 2jl to 29\% in SR 6jt. In SR 2jl the loose selection minimizes theoretical uncertainties and the impact of statistical fluctuations in the CRs, while the opposite is true in SR 6jt. 

For the backgrounds estimated with MC simulation-derived extrapolation factors, the primary common sources of systematic uncertainty are the jet energy scale (JES) calibration, jet energy resolution (JER), theoretical uncertainties, and limited event yields in the MC samples and data CRs. Correlations between uncertainties (for instance between JES or JER uncertainties in CRs and SRs) are taken into account where appropriate. 

The JES uncertainty was measured using the techniques
described in Refs.~\cite{Aad:2011he,Aad:2012vm,JetCalibRun2}. %
The JER uncertainty is estimated using the methods discussed in Refs.~\cite{Aad:2012ag,JetCalibRun2}. %
An additional uncertainty in the modelling of energy not associated with reconstructed objects, used in the calculation of $\met$ and measured with unassociated charged tracks, is also included.
The combined JES, JER and $\met$ (Jet/$\met$) uncertainty ranges from $<$1\% of the expected background in 2-jet SRs to 5\% in SR 6jt. 

Uncertainties arising from theoretical modelling of background processes are evaluated by comparing samples produced with different MC generators. 
The $W/Z$+jets events generated with \sherpa\ are compared to events generated with MG5\_aMC@NLO at leading order and interfaced to the \pythia~8.186 parton shower model. Uncertainties in the modelling of top quark pair production are estimated by comparing \textsc{Powheg-Box} to a\mcatnlo~\cite{Frixione:2002ik}, and by accounting for different generator and radiation tunes.  Uncertainties associated with PDF modelling of top quark pair production are found to be negligible.  Uncertainties in diboson production due to PDF, renormalization, factorization and resummation scale uncertainties (estimated by increasing and decreasing the scales used in the MC generators by a factor of two) are accounted for by applying a uniform 50\% uncertainty in all SRs, and are the dominant source of uncertainty in SRs 2jl and 2jm.  
Uncertainties associated with the modelling of $Z$+jets production are largest in the SRs with tight selection cuts (up to 14\%). %
The statistical uncertainty arising from the use of MC samples is largest (8\%) in SR 6jt. 
The uncertainties arising from the data-driven correction procedure applied to events selected in the CR$\gamma$ region, described in Section~\ref{sec:background}, are included in Table~\ref{tab:BreakdownSysSRCompressed} under `CR$\gamma$ corr. factor' and reach a value of 4\% in most of the SRs.  
The impact of lepton reconstruction uncertainties, and of the uncertainties related to the $b$-tag/$b$-veto efficiency, on the overall background uncertainty are found to be negligible for all SRs. 
The total background uncertainties for all SRs, broken down into the main contributing sources, are summarized in Table~\ref{tab:BreakdownSysSRCompressed}. 

\begin{table}
\scriptsize
\caption[Breakdown of uncertainty on background estimates]{
Breakdown of the dominant systematic uncertainties in the background estimates.
The individual uncertainties can be correlated, and do not necessarily add in quadrature to 
the total background uncertainty. $\Delta\mu$ uncertainties are the result of the control region statistical uncertainties and the systematic uncertainties entering a specific control region. 
In brackets, uncertainties are given relative to the expected total background yield, also presented in the Table. Empty cells (indicated by a `-') correspond to uncertainties lower than 1 per mil. \label{tab:BreakdownSysSRCompressed}}
\begin{center}
\begin{tabular}{| lrrrrrrr |}
\hline
Channel  &  {\bf 2jl}  &  {\bf 2jm }  &  {\bf 2jt }  &  {\bf 4jt }  &  {\bf 5j }  &  {\bf 6jm }  &  {\bf 6jt } \\ \hline
Total bkg  &  $283$  &  $191$  &  $23$  &  $4.6$  &  $13.2$  &  $6.9$  &  $4.2$ \\
Total bkg unc.  &  $\pm 24$  [$8\%$]  &  $\pm 21$  [$11\%$]  &  $\pm 4$  [$17\%$]  &  $\pm 1.1$  [$24\%$]  &  $\pm 2.2$  [$17\%$]  &  $\pm 1.5$  [$22\%$]  &  $\pm 1.2$  [$29\%$] \\
\hline
MC statistics  &  --  &  $\pm 2.3$ [$1\%$]  &  $\pm 0.5$ [$2\%$]  &  $\pm 0.31$ [$7\%$]  &  $\pm 0.5$ [$4\%$]  &  $\pm 0.4$ [$6\%$]  &  $\pm 0.32$ [$8\%$] \\
$\Delta\mu_{Z\rm+jets}$  &  $\pm 7$ [$2\%$]  &  $\pm 6$ [$3\%$]  &  $\pm 2.5$ [$11\%$]  &  $\pm 0.7$ [$15\%$]  &  $\pm 1.0$ [$8\%$]  &  $\pm 0.8$ [$12\%$]  &  $\pm 0.7$ [$17\%$] \\
$\Delta\mu_{W\rm+jets}$  &  $\pm 10$ [$4\%$]  &  $\pm 8$ [$4\%$]  &  $\pm 1.2$ [$5\%$]  &  $\pm 0.5$ [$11\%$]  &  $\pm 1.1$ [$8\%$]  &  $\pm 0.7$ [$10\%$]  &  $\pm 0.5$ [$12\%$] \\
$\Delta\mu$\ Top  &  $\pm 1.8$ [$1\%$]  &  $\pm 2.0$ [$1\%$]  &  $\pm 0.23$ [$1\%$]  &  $\pm 0.26$ [$6\%$]  &  $\pm 0.4$ [$3\%$]  &  $\pm 0.24$ [$3\%$]  &  $\pm 0.22$ [$5\%$] \\
$\Delta\mu_{\rm Multi-jet}$  &  $\pm 0.05$ [$0\%$]  &  $\pm 0.09$ [$0\%$]  &  $\pm 0.1$ [$0\%$]  &  --  &  --  &  --  &  -- \\
CR$\gamma$ corr. factor   &  $\pm 11$ [$4\%$]  &  $\pm 7$ [$4\%$]  &  $\pm 1.0$ [$4\%$]  &  $\pm 0.17$ [$4\%$]  &  $\pm 0.4$ [$3\%$]  &  $\pm 0.21$ [$3\%$]  &  $\pm 0.15$ [$4\%$] \\
Theory $Z$  &  $\pm 8$ [$3\%$]  &  $\pm 4$ [$2\%$]  &  $\pm 2.4$ [$10\%$]  &  $\pm 0.6$ [$13\%$]  &  $\pm 0.6$ [$5\%$]  &  $\pm 0.5$ [$7\%$]  &  $\pm 0.6$ [$14\%$] \\
Theory $W$  &  $\pm 2.9$ [$1\%$]  &  $\pm 2.5$ [$1\%$]  &  $\pm 0.5$ [$2\%$]  &  $\pm 0.29$ [$6\%$]  &  $\pm 0.7$ [$5\%$]  &  $\pm 0.5$ [$7\%$]  &  $\pm 0.4$ [$10\%$] \\
Theory top   &  $\pm 2.1$ [$1\%$]  &  $\pm 2.1$ [$1\%$]  &  $\pm 0.28$ [$1\%$]  &  $\pm 0.12$ [$3\%$]  &  $\pm 0.8$ [$6\%$]  &  $\pm 0.4$ [$6\%$]  &  $\pm 0.13$ [$3\%$] \\
Theory diboson   &  $\pm 15$ [$5\%$]  &  $\pm 15$ [$8\%$]  &  $\pm 1.0$ [$4\%$]  &  --  &  $\pm 1.0$ [$8\%$]  &  --  &  -- \\
Jet/\met\   &  $\pm 0.7$ [$0\%$]  &  $\pm 0.6$ [$0\%$]  &  $\pm 0.09$ [$0\%$]  &  $\pm 0.1$ [$2\%$]  &  $\pm 0.4$ [$3\%$]  &  $\pm 0.21$ [$3\%$]  &  $\pm 0.19$ [$5\%$] \\
\hline
\end{tabular}
\end{center}
\end{table}

\section{Results, interpretation and limits}
\label{sec:result}

The number of events observed in the data and the number of SM events expected to enter each of the signal regions, determined using the background-only fit, are shown in Table~\ref{tab:p0_UL} and Figure~\ref{fig:PlotSR}. The pre-fit background expectations are also shown in Table~\ref{tab:p0_UL} for comparison. The normalisation factors extracted simultaneously through the fit range for the different signal regions between 0.7 and 1.2 for $W$+jets, 0.4 and 0.8 for $\ttbar$(+EW) + single top, and 1.0 and 1.6 for  $Z/\gamma^*$+jets backgrounds.

\begin{table}
\scriptsize
\caption[p0 and UL]{Numbers of events observed in the signal regions used in the analysis compared with background expectations obtained from the fits described in the text. No signal contribution is considered in the CRs for the fit. Empty cells (indicated by a `-') correspond to estimates lower than $0.01$.
The p-values ($p_{0}$) give the probabilities of the observations being consistent with the estimated backgrounds. For an observed number of events lower than expected, the p-value is truncated at 0.5. Between parentheses, $p$-values are also given as the number of equivalent Gaussian standard deviations (Z).
Also shown are 95\% CL upper limits on the visible cross-section ($\langle\epsilon\sigma\rangle_{\rm obs}^{95}$), 
the visible number of signal events ($S_{\rm obs}^{95}$ ) and the number of signal events ($S_{\rm exp}^{95}$) 
given the expected number of background events (and $\pm 1\sigma$ excursions of the expectation). %
\label{tab:p0_UL}}
\begin{center}
\begin{tabular}{| lrrrrrrr | }
\hline
Signal Region & {\bf 2jl } & {\bf 2jm } & {\bf 2jt } & {\bf 4jt } & {\bf 5j } & {\bf 6jm } & {\bf 6jt } \\
\hline
\multicolumn{8}{|c|}{MC expected events} \\ \hline
Diboson &  $31$               &  $31$               &  $3.5$               &  $0.6$               &  $2.1$               &  $0.9$               &  $0.4$               \\
$Z/\gamma^*$+jets &  $167$               &  $104$               &  $13$               &  $2.0$               &  $5.4$               &  $2.8$               &  $1.4$               \\
$W$+jets &  $80$               &  $46$               &  $5.0$               &  $1.1$               &  $3.4$               &  $1.7$               &  $1.0$               \\
$\ttbar$(+EW) + single top &  $18$               &  $17$               &  $1.3$               &  $0.9$               &  $2.7$               &  $1.6$               &  $1.0$               \\
Multi-jet &  $0.7$               &  $0.8$               &  $0.04$               &  --               &  --               &  --               &  --               \\
\hline
Total MC &  $296$               &  $199$               &  $23$               &  $4.6$               &  $14$               &  $7.0$               &  $3.8$               \\
\hline
\multicolumn{8}{|c|}{Fitted background events} \\ \hline
Diboson & $31 \pm 15$ & $31 \pm 16$ & $3.5 \pm 1.8$ & $0.6 \pm 0.3$ & $2.1 \pm 1.1$ & $0.9 \pm 0.5$ & $0.43 \pm 0.27$ \\
$Z/\gamma^*$+jets & $170 \pm 16$ & $114 \pm 11$ & $16 \pm 4$ & $2.5 \pm 0.9$ & $6.0 \pm 1.3$ & $3.2 \pm 1.0$ & $2.2 \pm 1.0$ \\
$W$+jets & $68 \pm 10$ & $35 \pm 9$ & $3.5 \pm 1.3$ & $0.9 \pm 0.6$ & $3.5 \pm 1.3$ & $1.9 \pm 0.9$ & $1.2 \pm 0.7$ \\
$\ttbar$(+EW) + single top & $14 \pm 3$ & $10 \pm 3$ & $0.7 \pm 0.4$ & $0.6 \pm 0.3$ & $1.7 \pm 0.9$ & $0.9 \pm 0.5$ & $0.32 \pm 0.26$ \\
Multi-jet & $0.49 \pm 0.05$ & $0.6 \pm 0.4$ & $0.02 \pm 0.10$ & -- & -- & -- & -- \\
\hline
Total bkg & $283 \pm 24$ & $191 \pm 21$ & $23 \pm 4$ & $4.6 \pm 1.1$ & $13.2 \pm 2.2$ & $6.9 \pm 1.5$ & $4.2 \pm 1.2$ \\
\hline
Observed &  $263$                     &  $191$                     &  $26$                     &  $7$                     &  $7$                     &  $4$                     &  $3$                     \\
\hline
\hline
$\langle\epsilon{\rm \sigma}\rangle_{\rm obs}^{95}$ [fb]   &  $16$ & $15$   &  $5.2$   & $2.7$ &  $1.7$ & $1.7$ & $1.6$ \\
$S_{\rm obs}^{95}$     & $44$ &   $48$ &  $17$   & $8.7$ &  $5.4$  &  $5.4$ &  $5.0$ \\ 
$S_{\rm exp}^{95}$     & ${54}^{+21}_{-14}$ & $ { 48 }^{ +16 }_{ -10 }$  &  $ { 14.0 }^{ +5.4 }_{ -3.9 }$  & $ { 6.3 }^{ +2.9 }_{ -1.7 }$ &  $ { 8.7 }^{ +4.2 }_{ -1.9 }$ & $ { 6.6 }^{ +3.2 }_{ -1.5 }$  & $ { 5.7 }^{ +2.8 }_{ -1.5 }$ \\
$p_{0}$ ($\rm Z$)        & $ 0.50$~$(0.00)$ &  $ 0.50$~$(0.00)$ &  $ 0.40$~$(0.26)$  & $ 0.17$~$(0.94)$  & $ 0.50$~$(0.00)$ & $ 0.50$~$(0.00)$ & $ 0.50$~$(0.00)$ \\
\hline 
\end{tabular}

\vspace{0.5cm}
\end{center}
\end{table}

\begin{figure}[H]
\begin{center}
\includegraphics[width=0.7\textwidth]{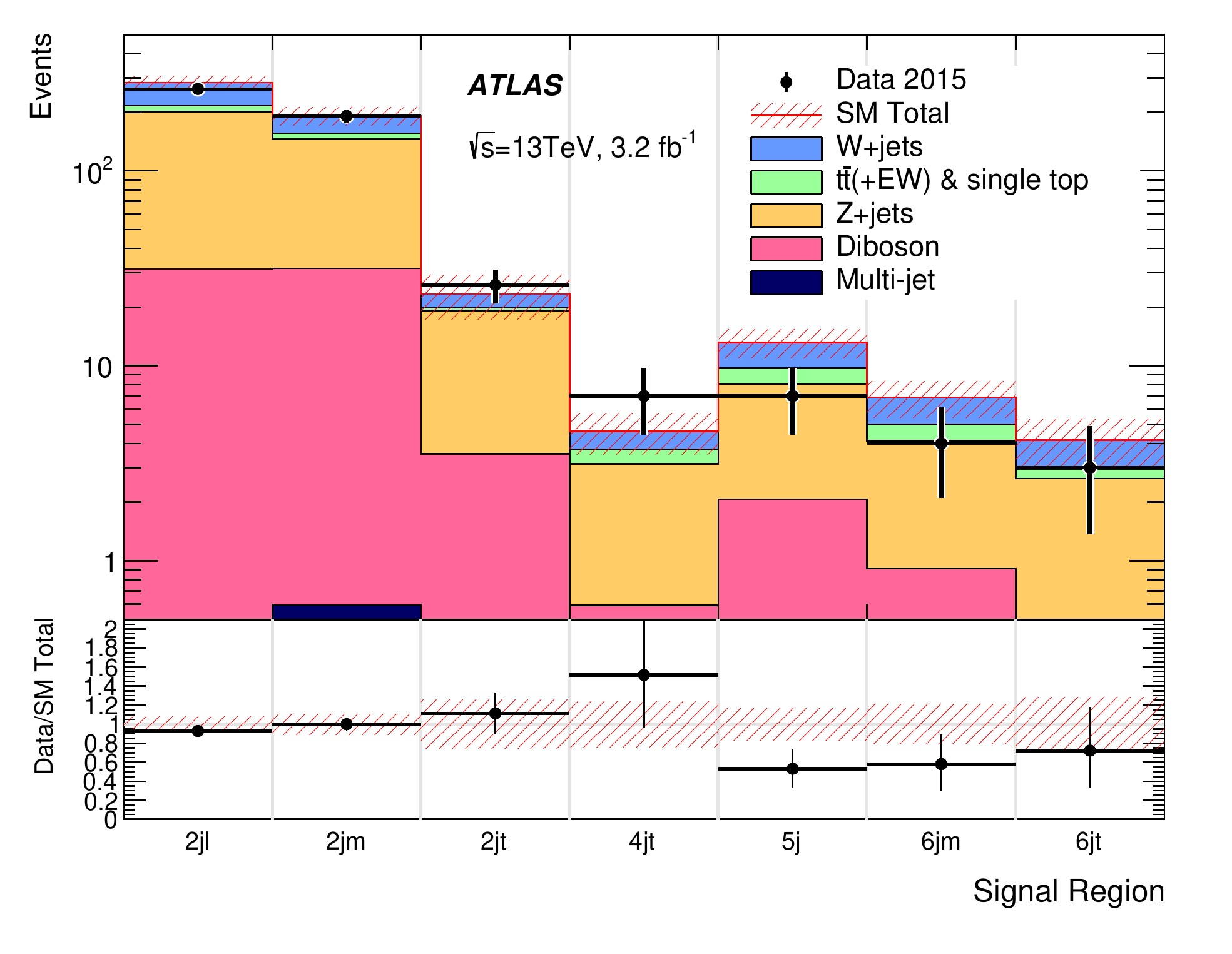}
\end{center}
\caption{\label{fig:PlotSR} Comparison of the observed and expected event yields as a function of signal region. The background expectations are those obtained from the background-only fits, presented in Table~\ref{tab:p0_UL}.}
\end{figure}

Distributions of $\meff({\rm incl.})$ obtained before the final selections on this quantity (but after applying all other selections), for data and the different MC samples normalized with the theoretical cross-sections, i.e. before applying the normalization from the CR fit, are shown in Figures~\ref{fig:sr2j}--\ref{fig:sr6j}. Examples of typical expected SUSY signals are shown for illustration. These signals correspond to the processes to which each SR is primarily sensitive -- $\squark\squark$ production for the lower jet-multiplicity SRs and $\gluino\gluino$ production for the higher jet-multiplicity SRs. In these figures, data and background distributions largely agree within uncertainties. The differences seen in the lower regions of $\meff({\rm incl.})$ distribution (1.2 -- 2.0 \TeV) in Figure~\ref{fig:sr6j} do not affect the background expectations in the signal regions since the backgrounds are normalized using control regions (Table~\ref{tab:crdefs}) with the same $\meff({\rm incl.})$ selections. 
The fit to the CRs for each SR compensates for the differences related to the overall normalization of the background seen in Figures~\ref{fig:sr2j}--\ref{fig:sr6j}, leading to the good agreements between data and post-fit expectations in the SRs observed in Table~\ref{tab:p0_UL} and Figure~\ref{fig:PlotSR}.

\begin{figure}[H]
\begin{center}
\subfigure[]{\includegraphics[width=0.42\textwidth]{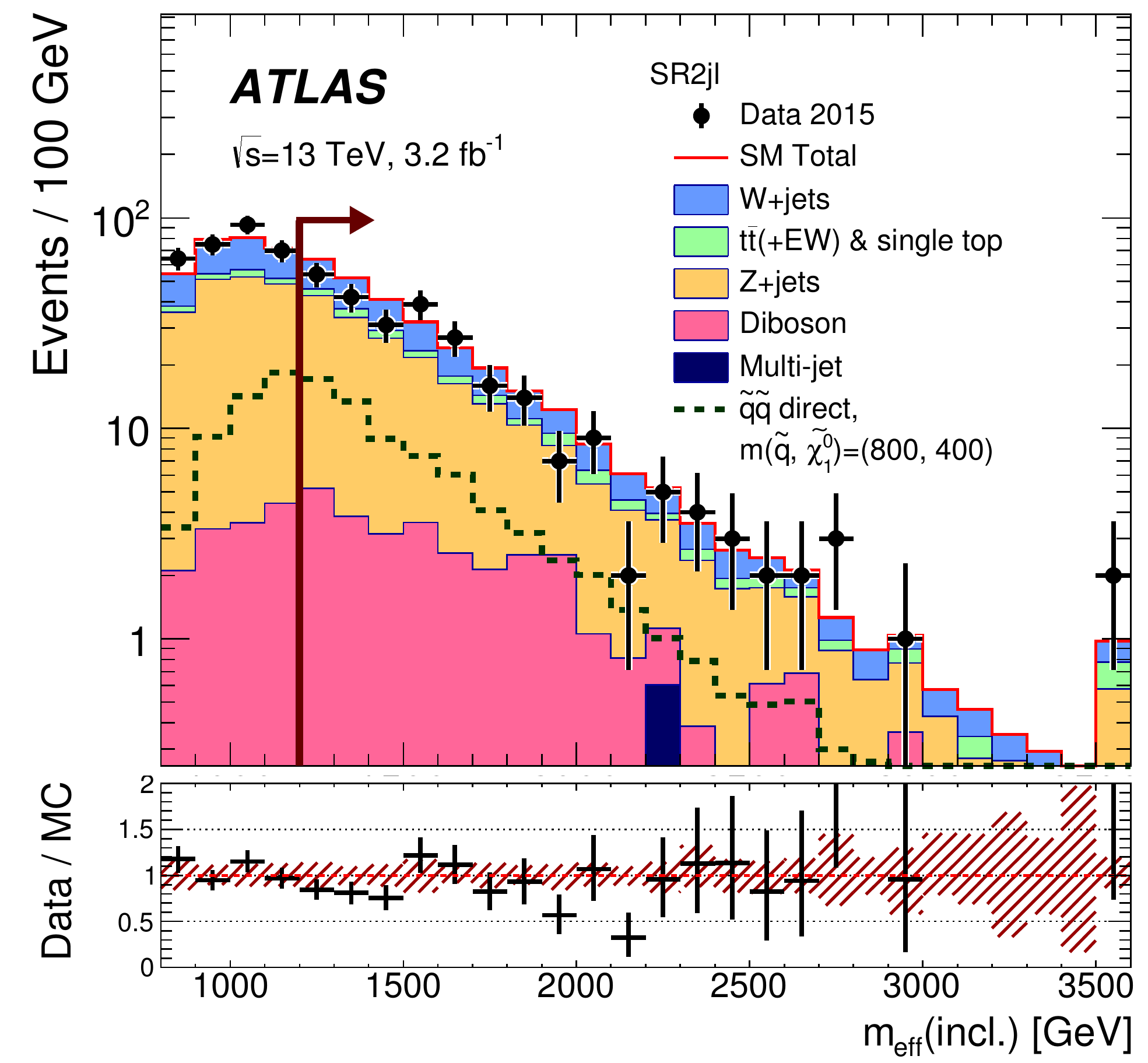}}
\subfigure[]{\includegraphics[width=0.42\textwidth]{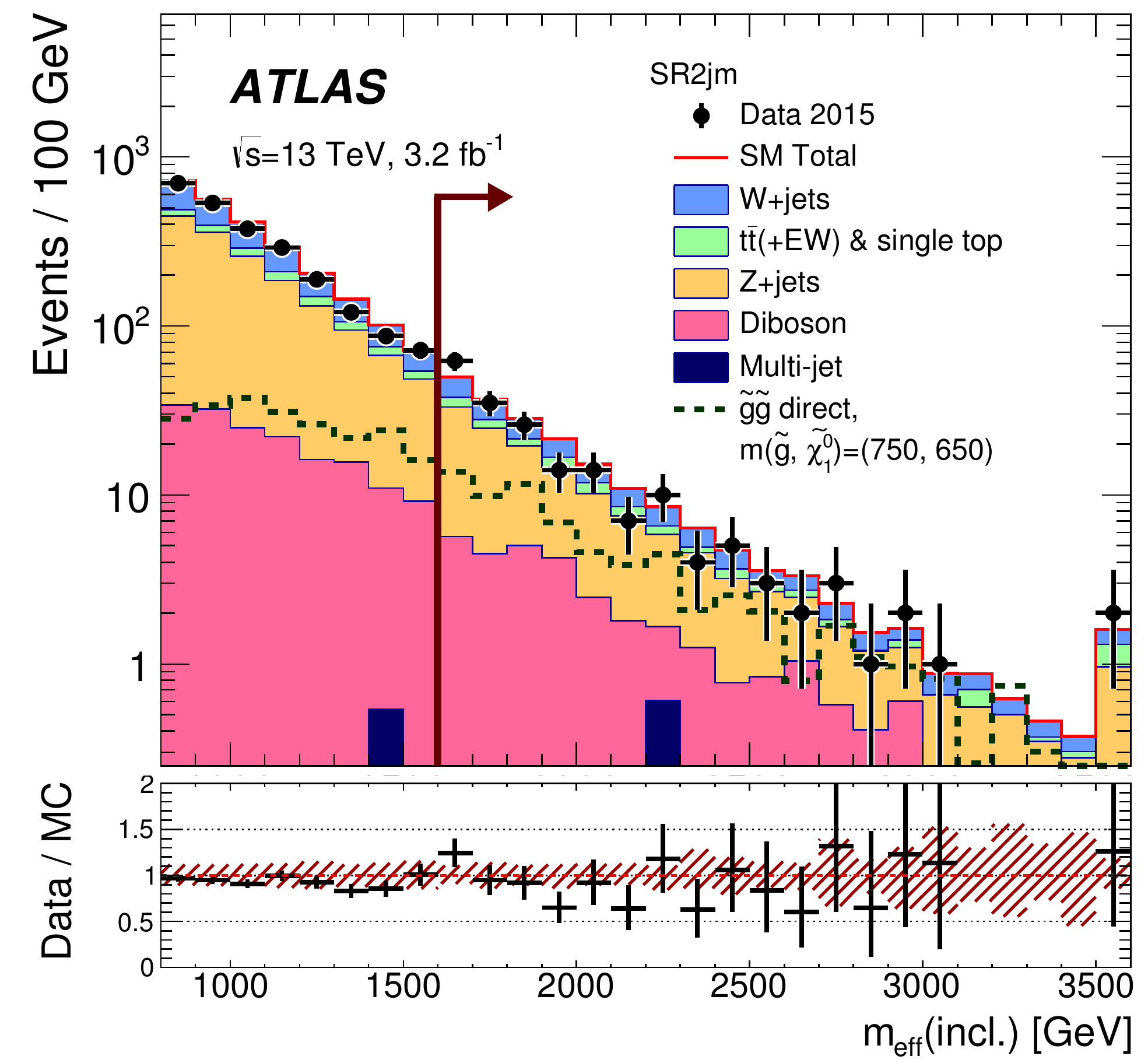}}\\
\subfigure[]{\includegraphics[width=0.42\textwidth]{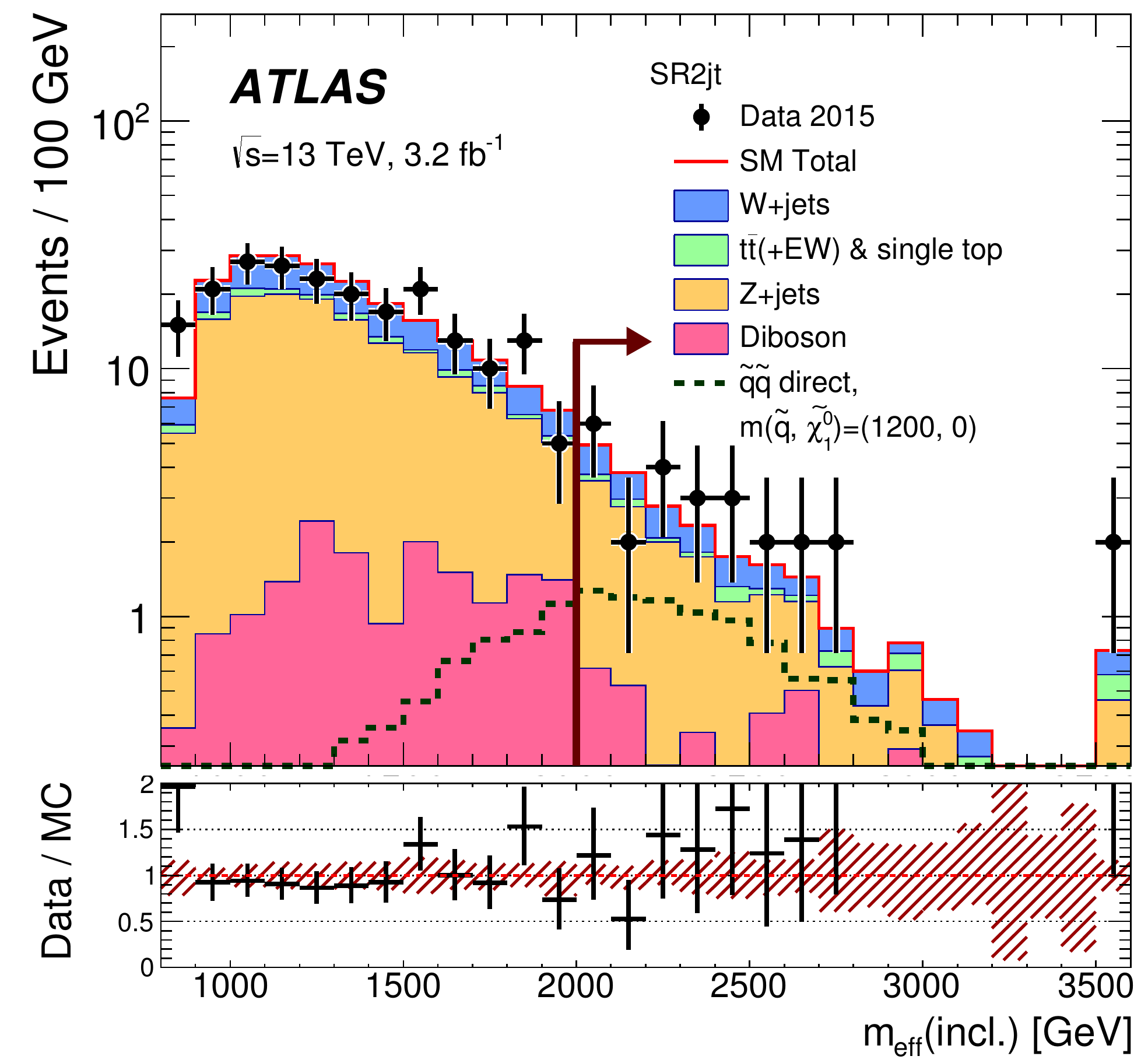}}
\end{center}
\caption{\label{fig:sr2j}Observed $\meff({\rm incl.})$ distributions for the (a) 2jl, (b) 2jm, (c) 2jt signal regions. The histograms denote the MC background expectations prior to the fits described in the text, normalized to cross-section times integrated luminosity. The last bin includes the overflow. In the lower panels the hatched (red) error bands denote the combined experimental, MC statistical and theoretical modelling uncertainties. The arrows indicate the values at which the requirements on $\meff({\rm incl.})$ are applied. 
Expected distributions for benchmark model points, normalized to NLO+NLL cross-section (Section~\ref{sec:montecarlo}) times integrated luminosity, are also shown for comparison (masses in \GeV). 
}
\end{figure}

\begin{figure}[H]
\begin{center}
\subfigure[]{\includegraphics[width=0.42\textwidth]{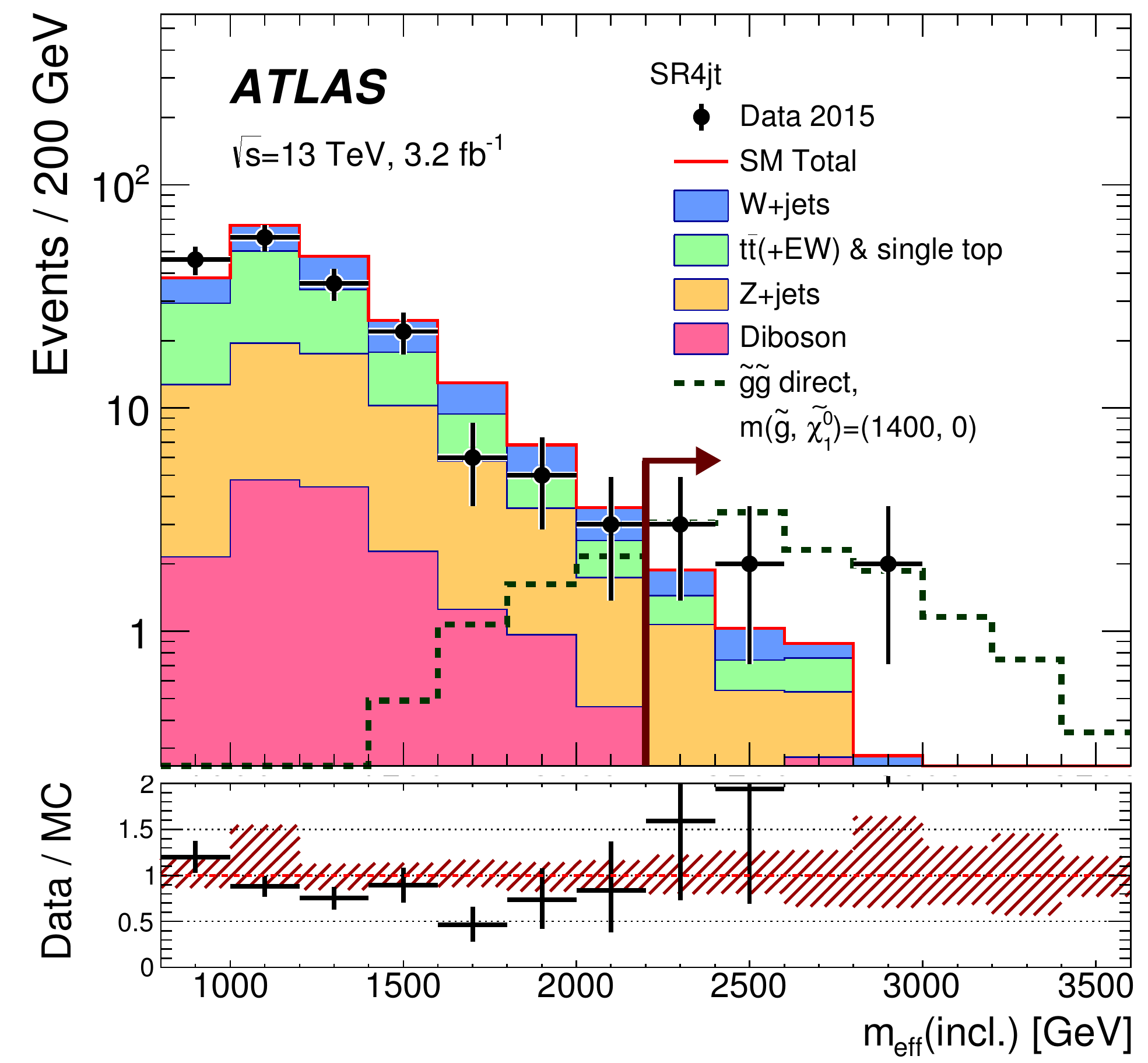}}
\subfigure[]{\includegraphics[width=0.42\textwidth]{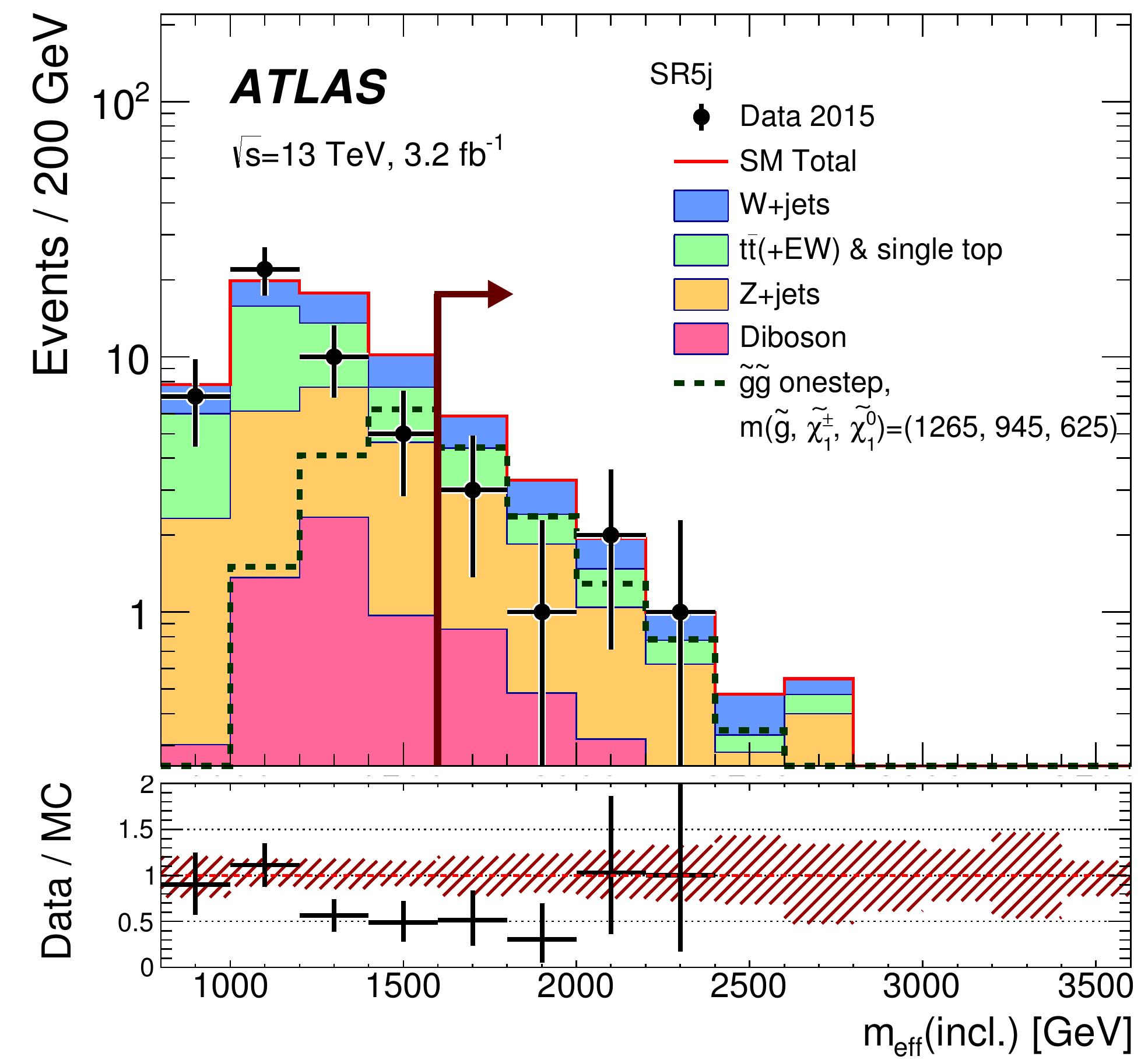}}\\
\subfigure[]{\includegraphics[width=0.42\textwidth]{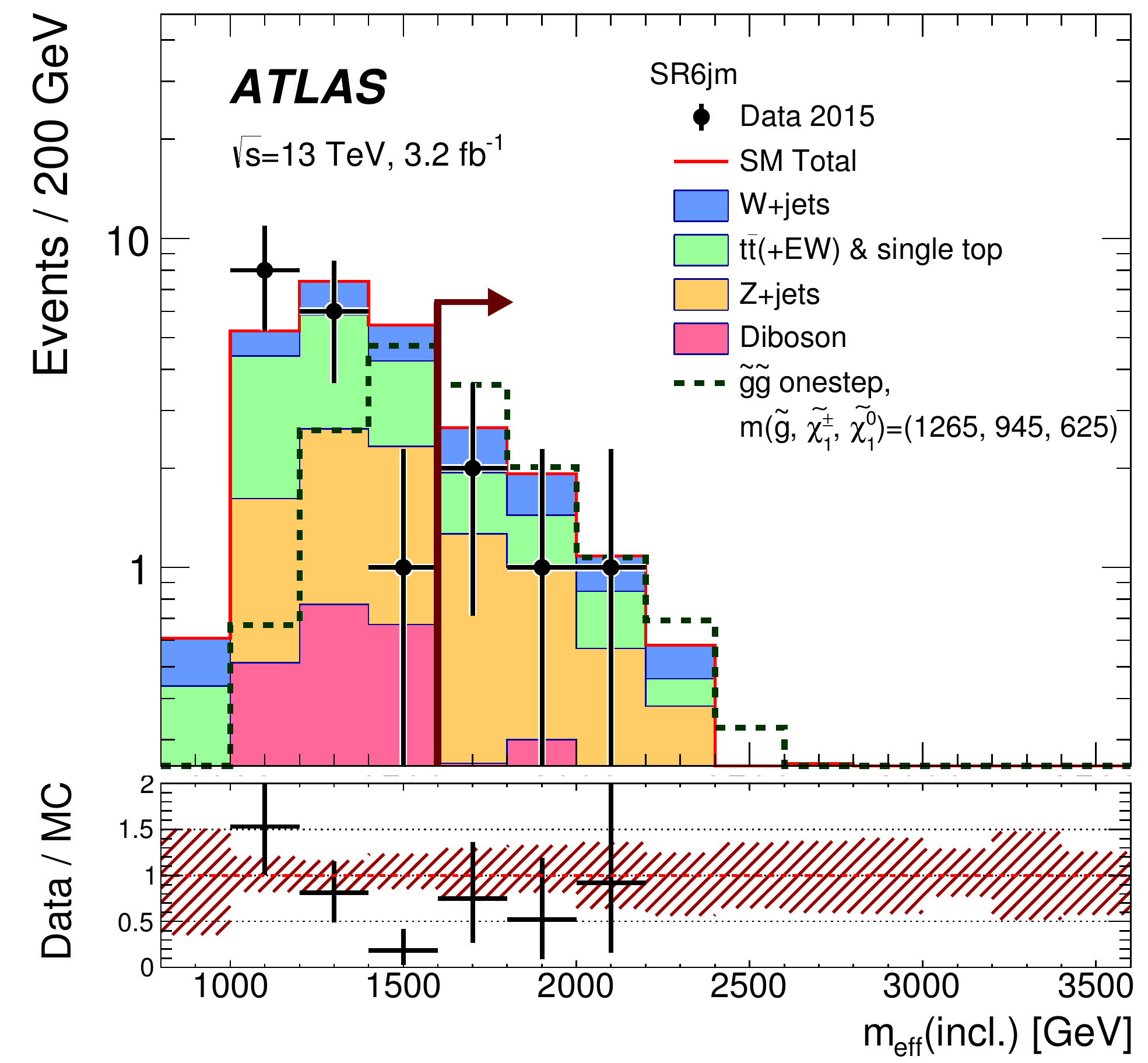}}
\subfigure[]{\includegraphics[width=0.42\textwidth]{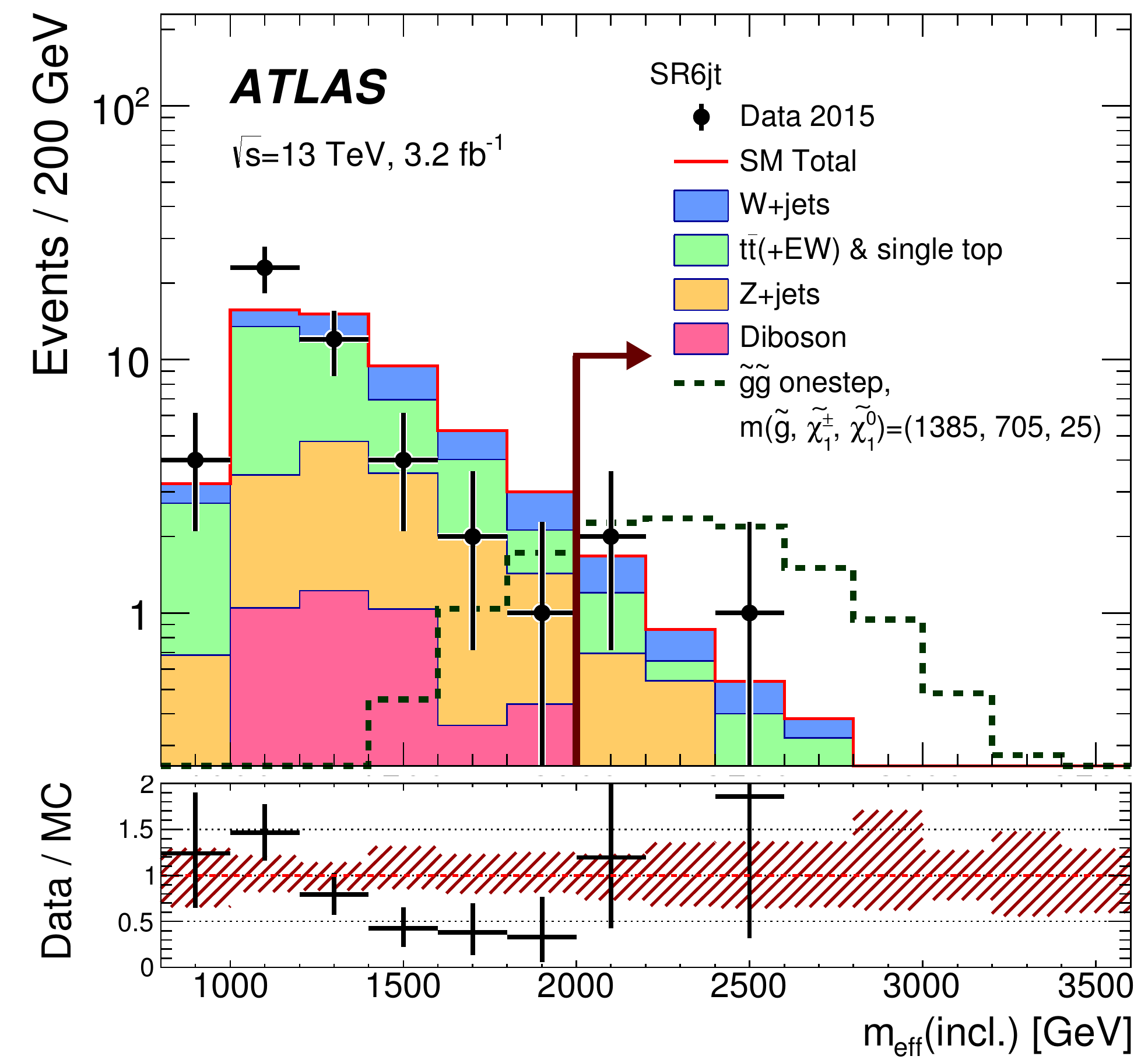}}
\end{center}
\caption{\label{fig:sr6j}Observed $\meff({\rm incl.})$ distributions for the (a) 4jt, (b) 5j, (c) 6jm and (d) 6jt signal regions. The histograms denote the MC background expectations prior to the fits described in the text, normalized to cross-section times integrated luminosity. The last bin includes the overflow. In the lower panels the hatched (red) error bands denote the combined experimental, MC statistical and theoretical modelling uncertainties. The arrows indicate the values at which the requirements on $\meff({\rm incl.})$ are applied.  
Expected distributions for benchmark model points, normalized to NLO+NLL cross-section (Section~\ref{sec:montecarlo}) times integrated luminosity, are also shown for comparison (masses in \GeV).  
}
\end{figure}

In the absence of a statistically significant excess, limits are set on contributions to the SRs from BSM physics. Upper limits at 95\% CL on the number of BSM signal events in each SR and the corresponding visible BSM cross-section are derived from the model-independent fits described in Section~\ref{sec:strategy} using the $CL_{\rm s}$ prescription. Limits are evaluated using MC pseudo-experiments. 
The results are presented in Table~\ref{tab:p0_UL}. 

The model-dependent fits in all the SRs are then used to set limits on specific classes of SUSY models, using the result from the SR with the best expected sensitivity at each point in each model parameter space. %
`Observed limits' are calculated from the observed SR event yields for the nominal signal cross-section. %
`Expected limits' are calculated by setting the nominal event yield in each SR to the corresponding mean expected background.

In Figure~\ref{fig:directLimit}, limits are shown for two classes of simplified models in which only direct production of light-flavour squark or gluino pairs are considered. 
In these simplified model scenarios, the upper limit of the excluded light-flavour squark mass region is 
1.03~\TeV\ assuming massless $\ninoone$, as obtained from the signal region 2jt. 
The corresponding limit on the gluino mass is 1.51~\TeV\, if the $\ninoone$ is massless, as obtained from the signal region 4jt. The best sensitivity in the region of parameter space where the mass difference between the squark (gluino) and the lightest neutralino is small is obtained from the signal region 2jm.

\begin{figure}[H]
\begin{center}
\subfigure[]{\includegraphics[width=0.60\textwidth]{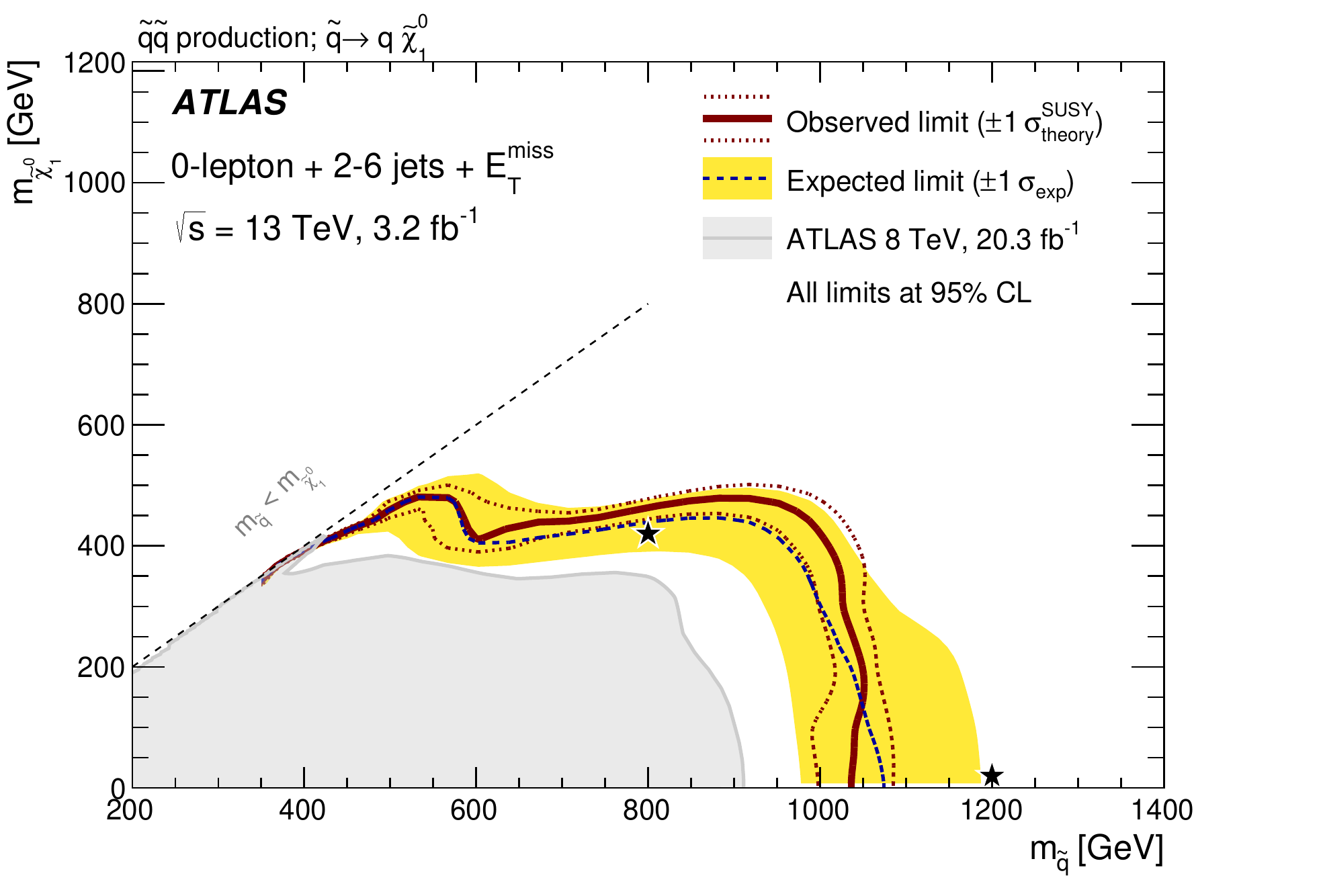}}
\subfigure[]{\includegraphics[width=0.60\textwidth]{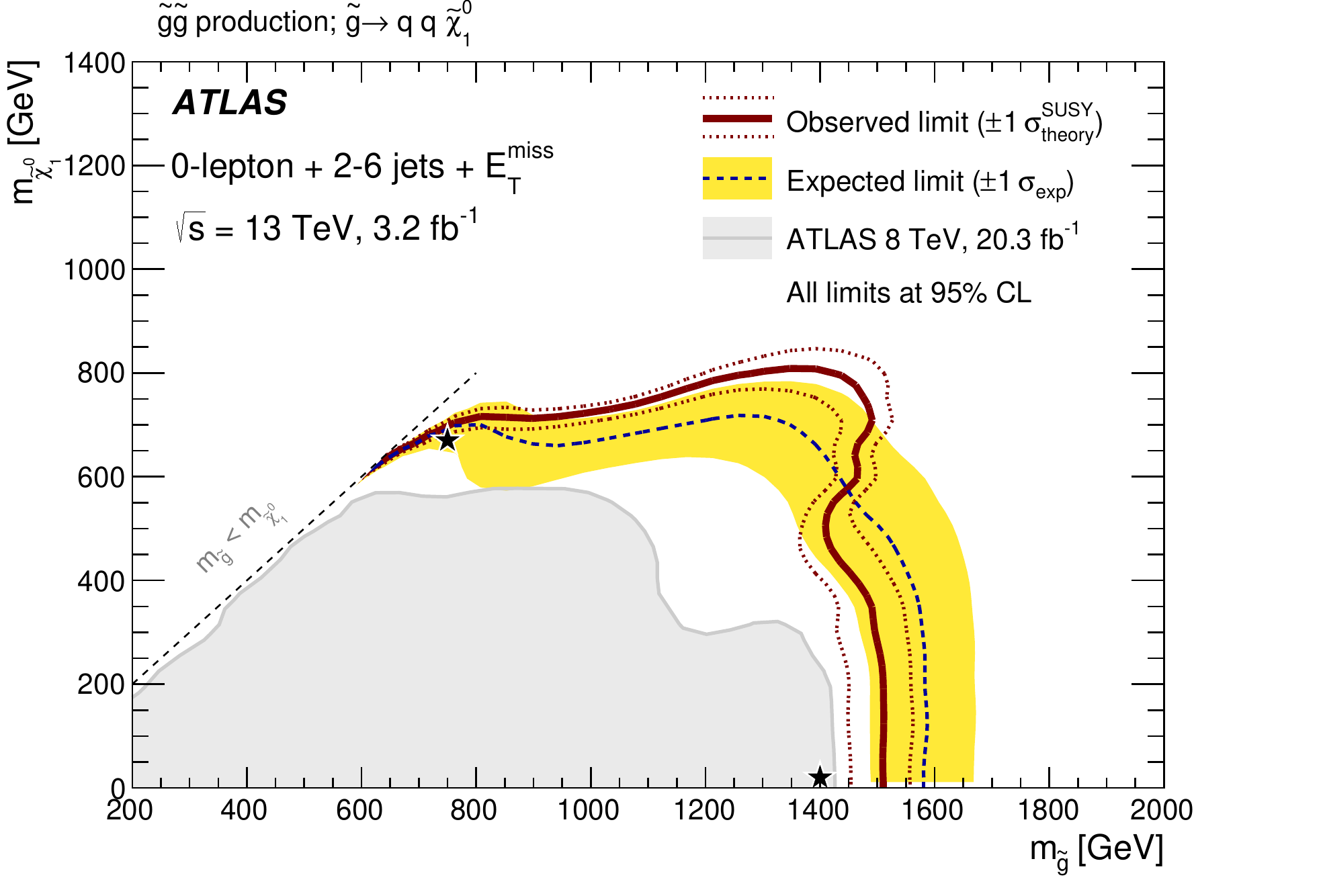}}
 \caption{ Exclusion limits for direct production of (a) light-flavour squark pairs with decoupled gluinos and (b) gluino pairs with decoupled squarks. Gluinos (light-flavour squarks) are required to decay to two quarks (one quark) and a neutralino LSP.  Exclusion limits are obtained by using the signal region with the best expected sensitivity at each point.  
The blue dashed lines show the expected limits at 95\% CL, with the light (yellow) bands indicating the $1\sigma$ excursions due to experimental and background-only  theoretical uncertainties.
Observed limits are indicated by medium dark (maroon) curves where the solid contour represents the nominal limit, and the dotted lines are obtained by varying the signal cross-section by the renormalization and factorization scale and PDF uncertainties. 
Results are compared with the observed limits obtained by the previous ATLAS search~\cite{summaryPaper}. %
The black stars indicate the benchmark models used in Figures~\ref{fig:sr2j} and \ref{fig:sr6j}. 
\label{fig:directLimit}}
\end{center}
\end{figure}

In Figure~\ref{fig:limitSMonestep}, limits are shown for pair-produced gluinos each decaying via an intermediate $\chinoonepm$ to two quarks, a $W$ boson and a $\ninoone$. Results are presented for simplified models in which the mass of the chargino $\chinoonepm$ is fixed to $m(\chinoonepm)=(m(\gluino)+m(\ninoone))/2$. For a $\ninoone$ mass of $\sim$~200~\GeV, the lower limit on the gluino mass, obtained from the signal region 4jt, extends up to 1.5~\TeV\ in this model.  In the region of parameter space where the mass difference between the gluino and the lightest neutralino is small, the best sensitivity is obtained from the signal region 2jm. Results are compared with the observed limits obtained from the statistical combination of the search with no lepton and the search with one isolated lepton, high-$\pt$ jets and missing transverse momentum performed at ATLAS~\cite{summaryPaper} using the 8 \TeV\ data. Statistical combinations of these two searches, designed to be statistically independent in their signal and control region definitions, are performed in order to increase the exclusion reach in models in which at least two analyses obtain comparable sensitivities, and still provide the strongest exclusion limits in the region of parameter space in which the mass of gluino is between 700 and 1100 \GeV\, and the $\ninoone$ mass is above $\sim$~500~\GeV. 

\begin{figure}[H]
\begin{center}
\includegraphics[width=0.60\textwidth]{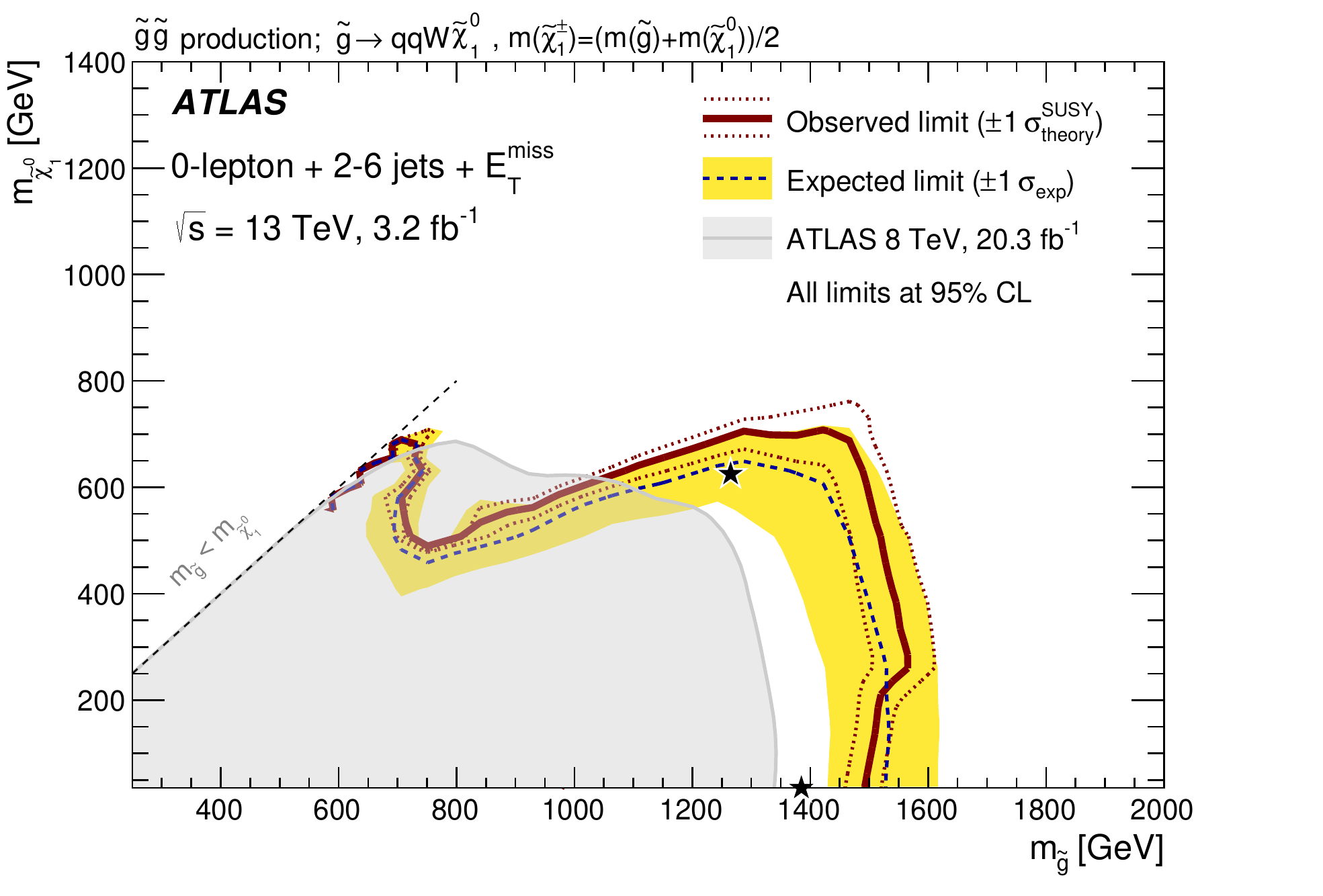}
\caption{Exclusion limits for pair-produced gluinos each decaying via an intermediate $\chinoonepm$ to two quarks, a $W$ boson and a $\ninoone$ for models with a fixed $m(\chinoonepm)=(m(\gluino)+m(\ninoone))/2$ and varying values of $m(\gluino)$ and $m({\ninoone})$. Exclusion limits are obtained by using the signal region with the best expected sensitivity at each point. The blue dashed lines show the expected limits at 95\% CL, with the light (yellow) bands indicating the $1\sigma$ excursions due to experimental and background-only theoretical uncertainties. Observed limits are indicated by medium dark (maroon) curves where the solid contour represents the nominal limit, and the dotted lines are obtained by varying the signal cross-section by the renormalization and factorization scale and PDF uncertainties. 
Results are compared with the observed limits obtained from the statistical combination of the search with no lepton and the search with one isolated lepton, high-$\pt$ jets and missing transverse momentum performed at ATLAS~\cite{summaryPaper}. %
The black stars indicate the benchmark models used in Figure~\ref{fig:sr6j}. }
\label{fig:limitSMonestep}
\end{center}
\end{figure}

\FloatBarrier
\section{Conclusion}
\label{sec:conclusion}
This paper reports a search for squarks and gluinos in final states containing
high-\pT{} jets, large missing transverse momentum but no electrons or muons, based on a 3.2~\ifb\ dataset of $\sqrt{s}=13 \TeV$ proton--proton collisions recorded by the ATLAS experiment at the LHC in 2015. Good agreement is seen between the numbers of events observed in the data and the numbers of events expected from SM processes. 

Results are interpreted in terms of simplified models with only light-flavour squarks, or gluinos, together with a neutralino LSP, with the masses of all the other SUSY particles set beyond the reach of the LHC.   
For a massless lightest neutralino, gluino masses below 1.51~\TeV\ are excluded at the 95\% confidence level in a simplified model with only gluinos and the lightest neutralino. 
For a simplified model involving the strong production of squarks of the first and second generations, with decays to a massless lightest neutralino, squark masses below 1.03~\TeV\ are excluded, assuming mass-degenerate squarks. 
In simplified models with pair-produced gluinos, each decaying via an intermediate $\chinoonepm$ to two quarks, a $W$ boson and a $\ninoone$, gluino masses below 1.5~\TeV\ are excluded for $\ninoone$ masses of $\sim$~200~\GeV.   
These results substantially extend the region of supersymmetric parameter space excluded by previous LHC searches.

\section*{Acknowledgements}

We thank CERN for the very successful operation of the LHC, as well as the
support staff from our institutions without whom ATLAS could not be
operated efficiently.

We acknowledge the support of ANPCyT, Argentina; YerPhI, Armenia; ARC, Australia; BMWFW and FWF, Austria; ANAS, Azerbaijan; SSTC, Belarus; CNPq and FAPESP, Brazil; NSERC, NRC and CFI, Canada; CERN; CONICYT, Chile; CAS, MOST and NSFC, China; COLCIENCIAS, Colombia; MSMT CR, MPO CR and VSC CR, Czech Republic; DNRF and DNSRC, Denmark; IN2P3-CNRS, CEA-DSM/IRFU, France; GNSF, Georgia; BMBF, HGF, and MPG, Germany; GSRT, Greece; RGC, Hong Kong SAR, China; ISF, I-CORE and Benoziyo Center, Israel; INFN, Italy; MEXT and JSPS, Japan; CNRST, Morocco; FOM and NWO, Netherlands; RCN, Norway; MNiSW and NCN, Poland; FCT, Portugal; MNE/IFA, Romania; MES of Russia and NRC KI, Russian Federation; JINR; MESTD, Serbia; MSSR, Slovakia; ARRS and MIZ\v{S}, Slovenia; DST/NRF, South Africa; MINECO, Spain; SRC and Wallenberg Foundation, Sweden; SERI, SNSF and Cantons of Bern and Geneva, Switzerland; MOST, Taiwan; TAEK, Turkey; STFC, United Kingdom; DOE and NSF, United States of America. In addition, individual groups and members have received support from BCKDF, the Canada Council, CANARIE, CRC, Compute Canada, FQRNT, and the Ontario Innovation Trust, Canada; EPLANET, ERC, FP7, Horizon 2020 and Marie Sk{\l}odowska-Curie Actions, European Union; Investissements d'Avenir Labex and Idex, ANR, R{\'e}gion Auvergne and Fondation Partager le Savoir, France; DFG and AvH Foundation, Germany; Herakleitos, Thales and Aristeia programmes co-financed by EU-ESF and the Greek NSRF; BSF, GIF and Minerva, Israel; BRF, Norway; Generalitat de Catalunya, Generalitat Valenciana, Spain; the Royal Society and Leverhulme Trust, United Kingdom.

The crucial computing support from all WLCG partners is acknowledged
gratefully, in particular from CERN and the ATLAS Tier-1 facilities at
TRIUMF (Canada), NDGF (Denmark, Norway, Sweden), CC-IN2P3 (France),
KIT/GridKA (Germany), INFN-CNAF (Italy), NL-T1 (Netherlands), PIC (Spain),
ASGC (Taiwan), RAL (UK) and BNL (USA) and in the Tier-2 facilities
worldwide.

\clearpage

\printbibliography

\clearpage

\FloatBarrier
\appendix

\clearpage
\input{atlas_authlist}
\end{document}

%% file: atlas_authlist.tex

\begin{flushleft}
{\Large The ATLAS Collaboration}

\bigskip

M.~Aaboud$^{\rm 136d}$,
G.~Aad$^{\rm 87}$,
B.~Abbott$^{\rm 114}$,
J.~Abdallah$^{\rm 65}$,
O.~Abdinov$^{\rm 12}$,
B.~Abeloos$^{\rm 118}$,
R.~Aben$^{\rm 108}$,
O.S.~AbouZeid$^{\rm 138}$,
N.L.~Abraham$^{\rm 150}$,
H.~Abramowicz$^{\rm 154}$
% ...
\bigskip
\\
% ...
$^{12}$ Institute of Physics, Azerbaijan Academy of Sciences, Baku, Azerbaijan\\
% ...
$^{65}$ University of Iowa, Iowa City IA, United States of America\\
% ...
$^{87}$ CPPM, Aix-Marseille Universit{\'e} and CNRS/IN2P3, Marseille, France\\
% ...
$^{108}$ Nikhef National Institute for Subatomic Physics and University of Amsterdam, Amsterdam, Netherlands\\
% ...
$^{114}$ Homer L. Dodge Department of Physics and Astronomy, University of Oklahoma, Norman OK, United States of America\\
% ...
$^{118}$ LAL, Univ. Paris-Sud, CNRS/IN2P3, Universit{\'e} Paris-Saclay, Orsay, France\\
% ...
$^{136}$ $^{(a)}$ Facult{\'e} des Sciences Ain Chock, R{\'e}seau Universitaire de Physique des Hautes Energies - Universit{\'e} Hassan II, Casablanca; $^{(b)}$ Centre National de l'Energie des Sciences Techniques Nucleaires, Rabat; $^{(c)}$ Facult{\'e} des Sciences Semlalia, Universit{\'e} Cadi Ayyad, LPHEA-Marrakech; $^{(d)}$ Facult{\'e} des Sciences, Universit{\'e} Mohamed Premier and LPTPM, Oujda; $^{(e)}$ Facult{\'e} des sciences, Universit{\'e} Mohammed V, Rabat, Morocco\\
% ...
$^{138}$ Santa Cruz Institute for Particle Physics, University of California Santa Cruz, Santa Cruz CA, United States of America\\
% ...
$^{150}$ Department of Physics and Astronomy, University of Sussex, Brighton, United Kingdom\\
% ...
$^{154}$ Raymond and Beverly Sackler School of Physics and Astronomy, Tel Aviv University, Tel Aviv, Israel\\
% ...
$^{a}$ Also at Department of Physics, King's College London, London, United Kingdom\\
$^{b}$ Also at Institute of Physics, Azerbaijan Academy of Sciences, Baku, Azerbaijan\\
$^{c}$ Also at Novosibirsk State University, Novosibirsk, Russia\\
$^{d}$ Also at TRIUMF, Vancouver BC, Canada\\
$^{e}$ Also at Department of Physics {\&} Astronomy, University of Louisville, Louisville, KY, United States of America\\
% ...
\end{flushleft}
